\documentclass[]{imag-ms-template}

\usepackage{xcolor}
\usepackage{tabularx}
\usepackage{array}
\usepackage{multirow}
\usepackage{adjustbox}
\usepackage{float}
\usepackage{soul}

\title{Bayesian Insights into Exchange and Restriction in Gray Matter Diffusion MRI}

\author{Maëliss Jallais$^{1,2\ast}$, Quentin Uhl$^{3,4}$, Tommaso Pavan$^{3,4}$, Malwina Molendowska$^1$,\\
Derek K. Jones$^1$, Ileana Jelescu$^{3,4}$, Marco Palombo$^{1,2}$\\
{\small $^{1}$Cardiff University Brain Research Imaging Centre (CUBRIC), Cardiff University, Cardiff, United Kingdom,}\\
{\small $^{2}$School of Computer Science and Informatics, Cardiff University, Cardiff, United Kingdom,}\\
{\small $^{3}$Department of Radiology, Lausanne University Hospital (CHUV), Lausanne, Switzerland,}\\
{\small $^{4}$School of Biology and Medicine, University of Lausanne, Lausanne, Switzerland}\\
{\small $^\ast$Correspondence:  jallaism@cardiff.ac.uk}
}

\begin{document} 

\maketitle 

\keywords{Diffusion MRI, Gray Matter, Water Exchange, Bayesian Inference, Uncertainty, Degeneracy}

\begin{abstract}
Biophysical models in diffusion MRI (dMRI) hold promise for characterizing gray matter tissue microstructure. Yet, the reliability of their parameter estimates remains largely under-studied, especially in models that incorporate water exchange. In this study, we investigate the accuracy, precision, and presence of degeneracy of two recently proposed gray matter models, NEXI and SANDIX, using established acquisition protocols, on both simulated and \textit{in vivo} data. We employ µGUIDE, a Bayesian inference framework based on deep learning, to quantify parameter uncertainty and detect degeneracies, enabling a more interpretable assessment of model fits. Our results show that while some microstructural parameters, such as extra-cellular diffusivity and neurite signal fraction, are robustly estimated, others, including exchange time and soma radius, are often associated with high uncertainty and estimation bias, particularly under realistic noise conditions and reduced acquisition protocols. Comparison with non-linear least squares fitting highlights the critical advantage of uncertainty-aware methods: the ability to flag and filter out unreliable estimates. Together, these findings emphasize the need to report uncertainty and account for model degeneracies when interpreting model-based estimates. Our study advocates for the integration of probabilistic fitting approaches into imaging pipelines to improve reproducibility and biological interpretability.
\end{abstract}

\section{Introduction}

Diffusion-weighted magnetic resonance imaging (dMRI) is a promising technique for characterizing brain microstructure \textit{in vivo} using a paradigm called microstructure imaging \citep{alexander_imaging_2017,jelescu_challenges_2020,novikov_quantifying_2018}. By fitting biophysical models to dMRI signals, quantitative measures that reflect histologically meaningful features of tissue microstructure can be estimated.

Brain white matter (WM) microstructure has been extensively studied, and commonly-used biophysical models typically include two or three non-exchanging compartments under a Gaussian diffusion assumption \citep{panagiotaki_compartment_2012}: axons modeled as a collection of zero-radius cylinders (namely "sticks"), following an orientation dispersion function (ODF); an anisotropic extra-neurite space modeled by an ellipsoid aligned with the main stick bundle direction; and, when needed, cerebrospinal fluid (CSF) contamination, included as a free water component. Numerous implementations have been proposed, varying the shape of the ODF, assumptions about the extra-neurite space diffusion, or introducing simplifying assumptions between the compartmental diffusivities \citep{fieremans_white_2011,zhang_noddi:_2012,novikov_rotationally-invariant_2018,novikov_quantifying_2018,reisert_disentangling_2017,jespersen_neurite_2010}.

For a collection of impermeable sticks, the directionally-averaged signal $S$ is expected to follow a power-law decay for high b-values $b$, with $S \propto b^{-1/2}$ \citep{veraart_scaling_2019,lee_vivo_2020}. While WM approximately follows this behavior, deviations from this power law have been observed, reflecting sensitivity to finite axonal diameters \citep{veraart_noninvasive_2020} and structural disorder \citep{fieremans_vivo_2016}. These deviations are substantially higher in gray matter (GM) \citep{mckinnon_dependence_2017,veraart_noninvasive_2020}, indicating that GM microstructure cannot be captured by the same models.

This motivated the development of GM-specific biophysical models. Previous studies have explored multiple mechanisms underlying the distinct high-b behavior observed in GM \citep{jespersen_modeling_2007,jespersen_neurite_2010,komlosh_detection_2007,shemesh_microscopic_2011,shemesh_mapping_2012,truong_cortical_2014}. Three main hypotheses have been proposed to explain the stronger deviation from the stick power law. The first one attributes it to the presence of cell bodies (namely somas) \citep{palombo_sandi_2020,olesen_diffusion_2022,fang_diffusion_2020}, and adds an extra spherical compartment modeling restricted diffusion within somas, such as in the Soma And Neurite Density Imaging (SANDI) model \citep{palombo_sandi_2020}. Somas are typically neglected in WM due to their relatively small density (5-10\% \textit{ex vivo}) \citep{andersson_axon_2020,veraart_noninvasive_2020}, but occupy $\sim$10-20\% of gray matter by volume \citep{motta_dense_2019,shapson-coe_connectomic_2021}. Another approach suggests to account for exchange between the intra- and extra-neurite space across the neurite membranes, due to low myelination in GM tissue compared to WM, which allows water molecules to cross membrane boundaries and potentially bias measurements that assume impermeable compartments \citep{jelescu_neurite_2022,olesen_diffusion_2022}. Reported exchange times vary between 3-5 ms in \textit{ex vivo} rat cortex \citep{olesen_diffusion_2022}, 10-50 ms in \textit{in vivo} rat cortex \citep{jelescu_neurite_2022} and perfused neonatal mouse spinal cords \citep{williamson_magnetic_2019}, 10-80 ms in \textit{in vivo} human cortex \citep{veraart_noninvasive_2020,uhl_quantifying_2024,dong_vivo_2025,uhl_reduced_2025}, and up to 100-150 ms in astrocyte and neuron cultures \citep{yang_intracellular_2018}, rat brain \citep{quirk_equilibrium_2003} and rat brain cortical cultures \citep{bai_fast_2018}. Finally, non-Gaussian diffusion along the dendrites was suggested to break down as a result of structural disorder, such as neurite undulation, beading and dendritic spines \citep{ozarslan_influence_2018,lee_vivo_2020,henriques_microscopic_2019,chakwizira_role_2025,simsek_role_2025}.

Recent studies suggest that exchange is the primary mechanism underlying the diffusion time-dependence of the signal in both low- and high-b regimes, with the influence of soma becoming more significant at longer diffusion times within the high-b regime \citep{jelescu_neurite_2022,olesen_diffusion_2022}. To account for these effects, new biophysical models have been proposed, such as the Neurite Exchange Imaging (NEXI) model \citep{jelescu_neurite_2022} and the Standard Model with EXchange (SMEX) \citep{olesen_diffusion_2022}, which both incorporate water exchange between anisotropic compartments, and the Soma and Neurite Density Imaging with Exchange (SANDIX) model \citep{olesen_diffusion_2022, dong_vivo_2025}, which extends upon the SANDI model by accounting for both restriction in somas and permeative exchange across the neurites' membrane (i.e., axons and cell processes), thereby enabling estimation of soma size and fraction in addition to exchange-related parameters. 

Nevertheless, increasing model complexity comes at a cost. While models like SANDIX offer more accurate representations of GM microstructure, they introduce additional parameters and potential degeneracies in the parameter space, making them more difficult to fit robustly \citep{olesen_diffusion_2022,jallais_introducing_2024}. This challenge is further exacerbated by the choice of the acquisition protocols, where differences in diffusion time ($t$), gradient strength ($g$), and b-value sampling directly influence model sensitivity. Protocols used in preclinical imaging benefit from ultra-strong gradients (typically 300-1000 mT/m) and short diffusion times, enabling high sensitivity to microstructural features like soma and exchange. However, such protocols remain impractical in clinical settings due to time constraints and scanner hardware limitations (typically 20-80 mT/m gradients). Next-generation human MRI scanners equipped with ultra-strong gradient systems (up to 300 mT/m), such as the Connectom \citep{fan_investigating_2014,jones_microstructural_2018,huang_connectome_2021} or the MAGNUS \citep{abad_feasibility_2025} systems, bridge the gap between preclinical and clinical scanners by enabling acquisitions with short diffusion times and high b-values. While stronger gradients and shorter diffusion times increase sensitivity to microstructure, they do not fully resolve ambiguities between model parameters. Consequently, degeneracies, uncertainties, and trade-offs in parameter estimation must be systematically addressed to validate these models in both research and clinical settings \citep{afzali_spheriously_2021,jallais_introducing_2024}. While for WM models a proper analysis of model degeneracy has been done \citep{jelescu_degeneracy_2016,coelho_resolving_2019}, for complex GM models accounting for both restriction and exchange, this is still missing.

The goal of this study is to evaluate the fitting quality of SANDIX with both a preclinical and a Connectom acquisition protocol, by leveraging the µGUIDE Bayesian inference framework \citep{jallais_introducing_2024}. µGUIDE enables the estimation of posterior distributions, which enable to quantify the fitting quality via an uncertainty measure, and highlight parameter degeneracies. We compare the performance of SANDIX with the simpler model NEXI. Using simulations, we investigate how the acquisition protocol and noise characteristics influence uncertainty and the emergence of degeneracies. We fit both biophysical models to \textit{in vivo} human data of four healthy volunteers scanned on the high-gradient 3T Connectom scanner. Finally, we demonstrate the importance of incorporating uncertainty and degeneracy into the interpretation of results.

\section{Methods}

\subsection{Biophysical Models}
\label{sec:model}

\begin{figure}[htbp]
\begin{center}
\includegraphics[width=0.8\textwidth]{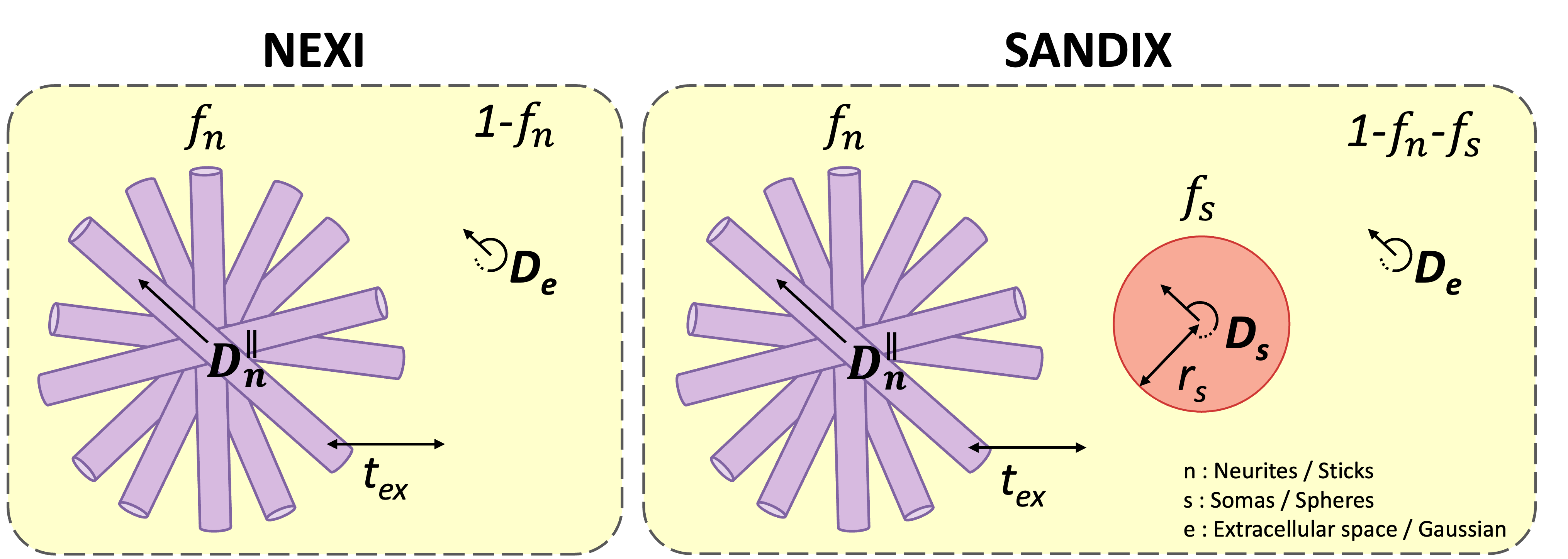}
\caption{Graphical representation of the considered gray matter biophysical models, with the following parameters: exchange time $t_{ex}$, neurites signal fraction $f_n$, parallel diffusivity $D_{n}^{\|}$, extra-cellular diffusivity $D_e$, soma signal fraction $f_s$ and soma radius $r_s$.}
\label{fig:graph_models}
\end{center}
\end{figure}

In this study, we focus on two biophysical models that explicitly account for permeative exchange between neurites and the extra-cellular space: NEXI \citep{jelescu_neurite_2022} and SANDIX \citep{olesen_diffusion_2022} (Figure~\ref{fig:graph_models}). While both models are rooted in the principles of the Kärger model \citep{fieremans_monte_2010,karger_nmr_1985}, they differ in two key aspects. First, NEXI is a two-compartment model that provides an analytical representation of the diffusion time-dependent signal under the narrow pulse approximation \citep{stejskal_use_1965,stejskal_restricted_1968}. In contrast, SANDIX extends on the three-compartment SANDI model by incorporating permeative exchange between neurites and extra-neurite compartments. This makes it more comprehensive in representing GM tissue microstructure, at the cost of increased fitting complexity. Additionally, while NEXI relies on the narrow pulse approximation, SANDIX is based on a generalization of the Kärger model to arbitrary gradient profiles, as described by \citet{ning_cumulant_2018}, which requires a numerical resolution of an ordinary differential equation (ODE).

\subsubsection{NEXI}

NEXI is a two-compartment biophysical model based on the anisotropic Kärger model for a coherent fiber tract. The Kärger model assumes that the exchange is barrier-limited,  i.e. $l_c^2\ll D_{n}^{\|} \cdot \tau_{i-e}$, where $l_c$ is  the characteristic size of  the intra-neurite compartment, $D_{n}^{\|}$ is the parallel diffusivity within neurites, and $\tau_{i-e}$ is the residence time within the intra-neurite compartment before exchanging to the extra-neurite one \citep{fieremans_monte_2010,karger_nmr_1985}. Neurites are modeled as a collection of isotropically oriented sticks, neglecting branching, finite processes and undulations \citep{palombo_sandi_2020,olesen_diffusion_2022}, with a signal fraction $f_n$. The extra-neurite space is modeled as isotropic Gaussian diffusion with diffusivity $D_e < D_{n}^{\|}$. Water exchange between these two compartments is captured by a characteristic exchange time $t_{ex}$.
The total diffusion signal is modeled as the spherical mean of kernel $\mathcal{K}$:
$$\bar{S}_{NEXI}(q, t; t_{ex}, D_{n}^{\|}, D_{e}, f_n)=\left.S\right|_{q=0} \cdot \int_0^1 \mathcal{K}(q, t, \mathbf{g} \cdot \mathbf{n} ; t_{ex}, D_{n}^{\|}, D_{e}, f_n) d(\mathbf{g} \cdot \mathbf{n})$$
with 
$$\mathcal{K}\left(q, t, \mathbf{g} \cdot \mathbf{n} ; t_{ex}, D_{n}^{\|}, D_{e}, f_n\right)=f^{\prime} e^{-q^2 t D_n^{\prime}}+\left(1-f^{\prime}\right) e^{-q^2 t D_e^{\prime}},$$

$$D_{n / e}^{\prime}=\frac{1}{2}\left\{D_n+D_e+\frac{1}{q^2 t_{e x}} \mp\left[\left[D_e-D_n+\frac{2 f_n-1}{q^2 t_{e x}}\right]^2+\frac{4 f_n(1-f_n)}{q^4 t_{e x}^2}\right]^{\frac{1}{2}}\right\},$$

$$f^{\prime}=\frac{1}{D_n^{\prime}-D_e^{\prime}}\left[f_n D_n+(1-f_n) D_e-D_e^{\prime}\right]$$

where $\mathbf{n}$ are the neurite orientations, $q$ is the wave vector along direction $\mathbf{g}$, and $D_n \equiv D_{n}^{\|}(\mathbf{g} \cdot \mathbf{n})^2$.

The four parameters of interest here are therefore $t_{ex}$, $D_{n}^{\|}$, $D_e$ and $f_n$.

\subsubsection{SANDIX}
\label{subsec:SANDIX}
\citet{olesen_diffusion_2022} proposed a generalization of NEXI to any gradient waveforms \citep{ning_cumulant_2018} dubbed Standard Model with EXchange (SMEX). The signal for a gradient direction $\mathbf{g}$ and intra-neurite signal fraction $f_n$ is defined as:
$$ S_{SMEX}(q, t, \mathbf{g} \cdot \mathbf{n}; t_{ex}, D_{n}^{\|}, D_{e}, f_n) = S_1(t; t_{ex}, D_{n}^{\|}, D_{e}, f_n) + S_2(t; t_{ex}, D_{n}^{\|}, D_{e}, f_n) ,$$

where $S_1(t; t_{ex}, D_{n}^{\|}, D_{e}, f_n)$ and $S_2(t; t_{ex}, D_{n}^{\|}, D_{e}, f_n)$ are obtained by integrating the following ODE:

$$
\frac{d}{d t}\left[\begin{array}{l}
S_1(t) \\
S_2(t)
\end{array}\right]=\left(\left[\begin{array}{cc}
- \frac{1-f_n}{t_{ex}} & \frac{f_n}{t_{ex}} \\
\frac{1-f_n}{t_{ex}} & -\frac{f_n}{t_{ex}}
\end{array}\right]-q^2(t)\left[\begin{array}{cc}
D_{n}^{\|} (\mathbf{g} \cdot \mathbf{n})^2 & 0 \\
0 & D_e
\end{array}\right]\right)\left[\begin{array}{l}
S_1(t) \\
S_2(t)
\end{array}\right],
$$
with $\mathbf{n}$ designing the orientation of a single stick, $q(t) \equiv \gamma \int_0^t g\left(t^{\prime}\right) d t^{\prime}$ and $g$ the magnitude of the diffusion gradient pulse. 

SANDIX adds a third compartment of impermeable spheres to SMEX to model soma. Spheres are modeled following the Gaussian Phase Approximation (GPA) \citep{balinov_nmr_1993,palombo_sandi_2020} with radius $r_s$, soma signal fraction $f_s$ and a fixed diffusivity $D_s = 3$ µm$^2$/ms:
$$
\begin{aligned}
\bar{S}_{sphere}\left(q, t; D_{s}, r_s\right) \approx & \ \exp \left\{-\frac{2(\gamma g)^2}{D_{s}} \sum_{m=1}^{\infty} \frac{\alpha_m^{-4}}{\alpha_m^2 r_s^2-2}\right. \\
& \left.\times\left[2 \delta-\frac{2+e^{-a_m^2 D_{s}(\Delta-\delta)}-2 e^{-a_m^2 D_{s} \delta}-2 e^{-\alpha_m^2 D_{s} \Delta}+e^{-\alpha_m^2 D_{s}(\Delta+\delta)}}{\alpha_m^2 D_{s}}\right]\right\},
\end{aligned}
$$
where $\delta$ and $\Delta$ are the diffusion pulse width and separation, and $\alpha_m$ the $m$th root of the equation $\left(\alpha r_s\right)^{-1} J_{\frac{3}{2}}\left(\alpha r_s\right)=J_{\frac{5}{2}}\left(\alpha r_s\right)$ with $J_{n}(x)$ the Bessel function of the first kind.

The exchange between somas and neurites is considered negligible at diffusion times below 20 ms \citep{palombo_sandi_2020}. Exchange between somas and extracellular water is also neglected due to negligible surface-to-volume ratio of somas relative to neurites and the small soma volume. 
The resulting direction-averaged signal is the following:
\begin{equation}
    \begin{split}
        \bar{S}_{SANDIX}(q, t; t_{ex}, D_{n}^{\|}, D_{e}, f_n, r_s, f_s)=& f_s \cdot \bar{S}_{sphere}(q, t; D_s, r_s) \\
        & + (1 - f_s) \cdot \bar{S}_{SMEX}(q, t; t_{ex}, D_{n}^{\|}, D_{e}, \frac{f_n}{1-f_s})
    \end{split}
\end{equation}
% $$\bar{S}_{SANDIX}(q, t; t_{ex}, D_{n}^{\|}, D_{e}, f, r_s, f_s)= f \cdot f_s \cdot \bar{S}_{sphere}(q, t; D_s, r_s) + (1 - f \cdot f_s) \cdot \bar{S}_{SMEX}(q, t; t_{ex}, D_{n}^{\|}, D_{e}, \frac{f \cdot (1-f_s)}{1-f \cdot f_s})$$

This model thus adds two additional parameters compared to NEXI, for a total of six parameters to estimate: $t_{ex}$, $D_{n}^{\|}$, $D_e$, $f_n$, $r_s$, $f_s$.

\subsection{Acquisition Protocols}
\label{sec:acq_prot}
We considered two Pulsed Gradient Spin Echo (PGSE) protocols from the literature, both previously used to fit the NEXI and SANDIX models described in Section~\ref{sec:model}. The first protocol is an extensive acquisition designed for \textit{ex vivo} experiments, originally introduced by \citet{olesen_diffusion_2022} to present SANDIX. The second protocol is the NEXI 3T Connectom protocol \citep{uhl_quantifying_2024}, designed to be compatible with \textit{in vivo} human acquisitions on the Connectom scanner. We  present simulation results for both protocols, as well as \textit{in vivo} results acquired with the NEXI 3T Connectom protocol (see Section \ref{subsec:dMRI_data} for details on data acquisition).

\subsubsection{Extensive \textit{ex vivo} Acquisition Protocol}
This acquisition protocol was designed to enable the detection of a stick power-law in GM and to explore the signal's dependence on diffusion time using a 16.4 T Bruker Aeon scanner with a Micro5 probe (yielding gradient amplitudes up to 3000 mT/m). The gradient pulse width was fixed to $\delta$ = 4.5 ms and three separation times were considered: $\Delta$ = {16, 11, 7.5} ms. For each diffusion time, low b-values were set to 0.1, 0.5, 1, 2, 3, 4, and 5 ms/µm$^2$. Higher b-values were sampled and approximately uniformly spaced in $b^{-1/2}$, as follows:
\begin{itemize}
    \item $\Delta$ = 16 ms: from 0.4 to 0.3 µm/ms$^{1/2}$ with a spacing of 0.025 µm/ms$^{1/2}$, and from 0.3 to 0.1 µm/ms$^{1/2}$ with a spacing of 0.0125 µm/ms$^{1/2}$, resulting in 21 b-values between 6.25 and 100 ms/µm$^2$.
    \item $\Delta$ = 11 ms: from 0.4 to 0.125 µm/ms$^{1/2}$ with a spacing of 0.025 µm/ms$^{1/2}$, resulting in 12 b-values between 6.25 and 64 ms/µm$^2$.
    \item $\Delta$ = 7.5 ms: from 0.4 to 0.15 µm/ms$^{1/2}$ with a spacing of 0.025 µm/ms$^{1/2}$, resulting in 11 b-values between 6.25 and 44.44 ms/µm$^2$.
\end{itemize}
Ten sampling directions were used for b-values inferior to 25 ms/µm$^2$, and thirty otherwise.

\subsubsection{NEXI 3T Connectom protocol}
\label{subsub:3_C_prot}
The NEXI 3T Connectom protocol was designed to estimate exchange time on a 3T Siemens Connectom scanner. This protocol consists of a combination of five b-values and four diffusion times, with b-values of 1, 2.5, 4, 6 and 7.5 ms/µm$^2$ with respectively 13, 25, 25, 32 and 65 directions, and $\Delta=$20, 29, 39 and 49 ms. The gradient pulse width was fixed at 9 ms.

\subsection{Evaluating Model Fit, Uncertainty, and Degeneracies using µGUIDE}

To quantitatively assess model fitting quality and identify potential parameter degeneracies, we used µGUIDE \citep{jallais_introducing_2024}, a general Bayesian inference framework based on simulation-based inference \citep{cranmer_frontier_2020} that allows to estimate posterior distributions of biophysical model parameters. 

µGUIDE is composed of two jointly optimized modules. First, a Multi-Layer Perceptron (MLP) is used to reduce the dimensionality of the input signal. Second, a Neural Posterior Estimator (NPE) approximates the posterior distribution by learning a conditional probability density estimator that minimizes the Kullback-Leibler divergence \citep{papamakarios_fast_2016}. The conditional probability density approximators used here belong to a class of neural networks called normalizing flows \citep{papamakarios_normalizing_2021}, a class of neural networks designed for flexible density estimation. A masked autoregressive flow architecture \citep{papamakarios_masked_2017,germain_made_2015} is implemented in µGUIDE. Further details on the training procedure are provided in Section~\ref{subsec:training}.

The posterior distributions produced by µGUIDE provide valuable insights into both the confidence of parameter estimates and the landscape of the solution space. In particular, multi-modal posteriors, which are characterized by multiple distinct peaks, indicate degeneracy, meaning that different parameter combinations can produce equally plausible fits to the observed signal. We extracted three key summary statistics from the estimated posterior distributions:
\begin{itemize}
    \item the maximum a posteriori (MAP) estimate, representing the most probable parameter configuration;
    \item an uncertainty measure, defined as the interquartile range of the 50\% most probable samples, which quantifies the dispersion of the posterior distribution. Lower uncertainty values indicate higher precision in the MAP estimates;
    \item the presence of degeneracy.
\end{itemize}

\subsection{Simulation Framework and Model Fitting Setup}
\label{subsec:training}
We generated synthetic dMRI signals across eight experimental conditions: both biophysical models (NEXI and SANDIX, see Section~\ref{sec:model}), each evaluated under the two acquisition protocols (see Section~\ref{sec:acq_prot}), with and without the addition of Rician noise.

For each scenario, parameter combinations were randomly sampled within biologically feasible ranges: $t_{ex}\in[1,150]$ ms; $f_n \in [0.05,0.95]$ for NEXI and $f_n \in [0.025,0.90]$ for SANDIX, $f_s \in [0.05,0.5]$; $D_{n}^{\|}$,$D_e\in[0.1,3.0]$ µm${}^2$/ms; and $r_s\in[1,30]$ µm. To ensure uniform sampling while enforcing the $D_e < D_{n}^{\|}$ constraint, we followed the transformation method from \citep{jallais_introducing_2024,jallais_inverting_2022}, using two independent variables $u_0$ and $u_1 \sim \mathcal{U}(0,1)$:
$$
\left\{\begin{array}{l}
D_{n}^{\|}=\sqrt{(3.0-0.1)^2 \cdot u_0}+0.1 \\
D_e=\left(D_{n}^{\|}-0.1\right) \cdot u_1+0.1
\end{array}\right.
$$

Training was performed using the uniform $u_0$ and $u_1$, and transformed back to $D_{n}^{\|}$ and $D_e$ after inference. All parameters are normalized within the µGUIDE framework for inference.

To mimic acquired signals, we added Rician noise with a median Signal-to-Noise Ratio (SNR) of 36, based on empirical noise distributions estimated using MP-PCA \citep{veraart_denoising_2016} on the Connectom acquisitions (see Sections~\ref{subsec:Connectom_acq}\&\ref{subsec:data_proc}). Both noiseless and noisy datasets were used to assess the impact of noise on parameter estimation and number of degeneracies.

µGUIDE was trained separately for each experimental condition using $7 \cdot 10^5$ simulations. Using larger sets did not provide significant improvements, while required more time to train. 5\% of the simulations were randomly selected for validation. µGUIDE's MLP module was used to reduce the input dimension to 14 features for NEXI, and 22 features for SANDIX. These numbers were selected as optimal after preliminary testing in which the number of MLP features was varied between 5 and 30. The network was trained using a learning rate of $10^{-3}$, a minibatch size of 128, and early stopping after 50 epochs if no improvement in validation loss was obtained. A fine-tuning phase followed, using a learning rate of $10^{-4}$ and starting from the best checkpoint.

For inference, parameters were estimated independently for each signal by drawing 50,000 samples from the learned conditional posterior via rejection sampling, following the default value in µGUIDE's implementation. This procedure enables recovery of the full posterior distribution for each model parameter. For each estimated posterior distribution, we estimated the MAP and uncertainty, and flagged degenerate posterior distributions.

\subsection{dMRI Data}
\label{subsec:dMRI_data}
\subsubsection{Data Acquisition}
\label{subsec:Connectom_acq}
We fitted NEXI and SANDIX on data from four healthy volunteers (2 of whom rescanned on a different day) scanned following the NEXI 3T Connectom protocol \citep{uhl_quantifying_2024} on a 3T Siemens Connectom MRI scanner with a gradient amplitude of 300 mT/m (Siemens Heathineers, Erlangen, Germany). Diffusion-weighted images were acquired following the sequence parameters presented in section \ref{subsub:3_C_prot}. 15 images at $b=0$ ms/µm$^2$ per $\Delta$ were also acquired. The other parameters were kept fixed: TE/TR=76/3700 ms, FOV=216 x 216 mm$^2$, 1.8-mm isotropic resolution, partial Fourier = 0.75, GRAPPA = 2 and multiband = 2. An MPRAGE was acquired for anatomical reference (1-mm isotropic resolution, FOV= 256 x 256 mm$^2$, 192 slices, TI/TR=857/2300 ms). The total scan time was 45 minutes. 

\subsubsection{Data Processing}
\label{subsec:data_proc}
Multi-shell multi-diffusion time data were preprocessed jointly. Pre-processing steps included MP-PCA magnitude denoising \citep{veraart_denoising_2016}, Gibbs ringing correction \citep{kellner_gibbs-ringing_2016}, and distortion and eddy current correction \citep{andersson_integrated_2016}. The cortical ribbon was segmented on the MPRAGE image using FastSurfer \citep{henschel_fastsurfer_2020} and projected onto the native diffusion image space using linear registration \citep{avants_advanced_2009}. Data were powder-averaged using the arithmetic mean and normalized by the mean $b=0$ ms/µm${}^2$ values before being fed to the trained µGUIDE for model parameters' inference.

\section{Results}

\subsection{Simulations}

\begin{figure}[htbp]
\begin{center}
\includegraphics[width=\textwidth]{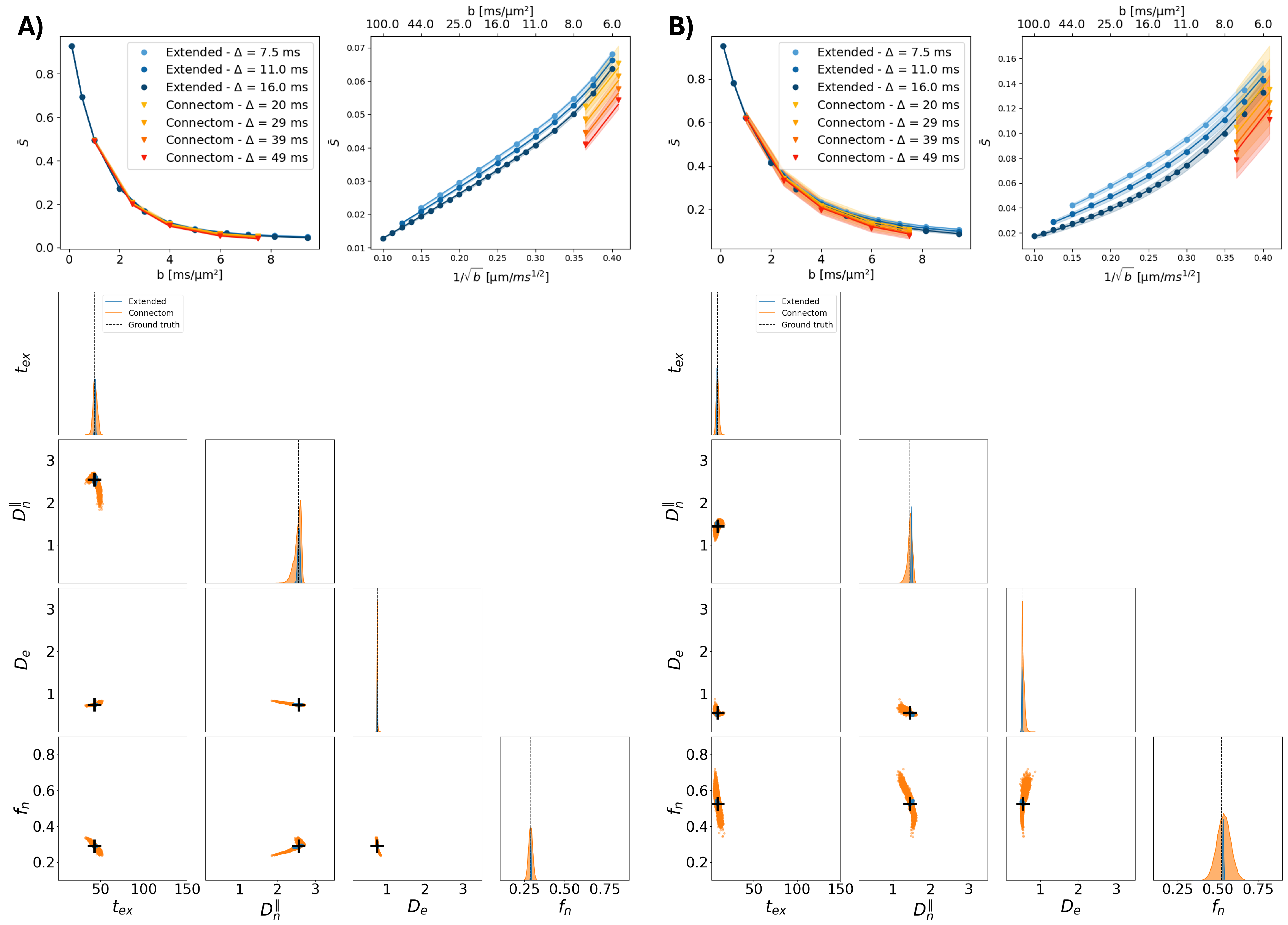 }
\caption{Simulated signals of the NEXI model, without noise, obtained from exemplar gray matter tissues using both acquisition protocols, and posterior distributions obtained with µGUIDE. For each subplot, upper rows: left plots show the signal with b-values inferior to 10 ms/µm$^2$ as a function of $b$, and right subplot shows the signal for b-values superior to 5 ms/µm$^2$ as a function of $b^{-1/2}$. Dots represent the simulated data, solid lines show signals generated from the MAP estimates, and shaded areas encompass all signals corresponding to parameter combinations sampled from the posterior distributions, illustrating their uncertainty. Bottom rows: Marginal posterior distributions of each parameter obtained with µGUIDE in the diagonal, and pairplots in the lower diagonal. Vertical black dashed lines and black crosses correspond to the ground truth values. Model parameters used to generate the signals are the following: A) $t_{ex}=43$ ms, $D_{n}^{\|}=2.55$ µm${}^2$/ms, $D_e=0.74$ µm${}^2$/ms, $f_n=0.29$ \citep{jelescu_neurite_2022}; B) $t_{ex}=8.15$ ms, $D_{n}^{\|}=1.45$ µm${}^2$/ms, $D_e=0.55$ µm${}^2$/ms, $f_n=0.525$ \citep{olesen_diffusion_2022}.}
\label{fig:signals_post_NEXI}
\end{center}
\end{figure}

\begin{figure}[htbp]
\begin{center}
\includegraphics[width=\textwidth]{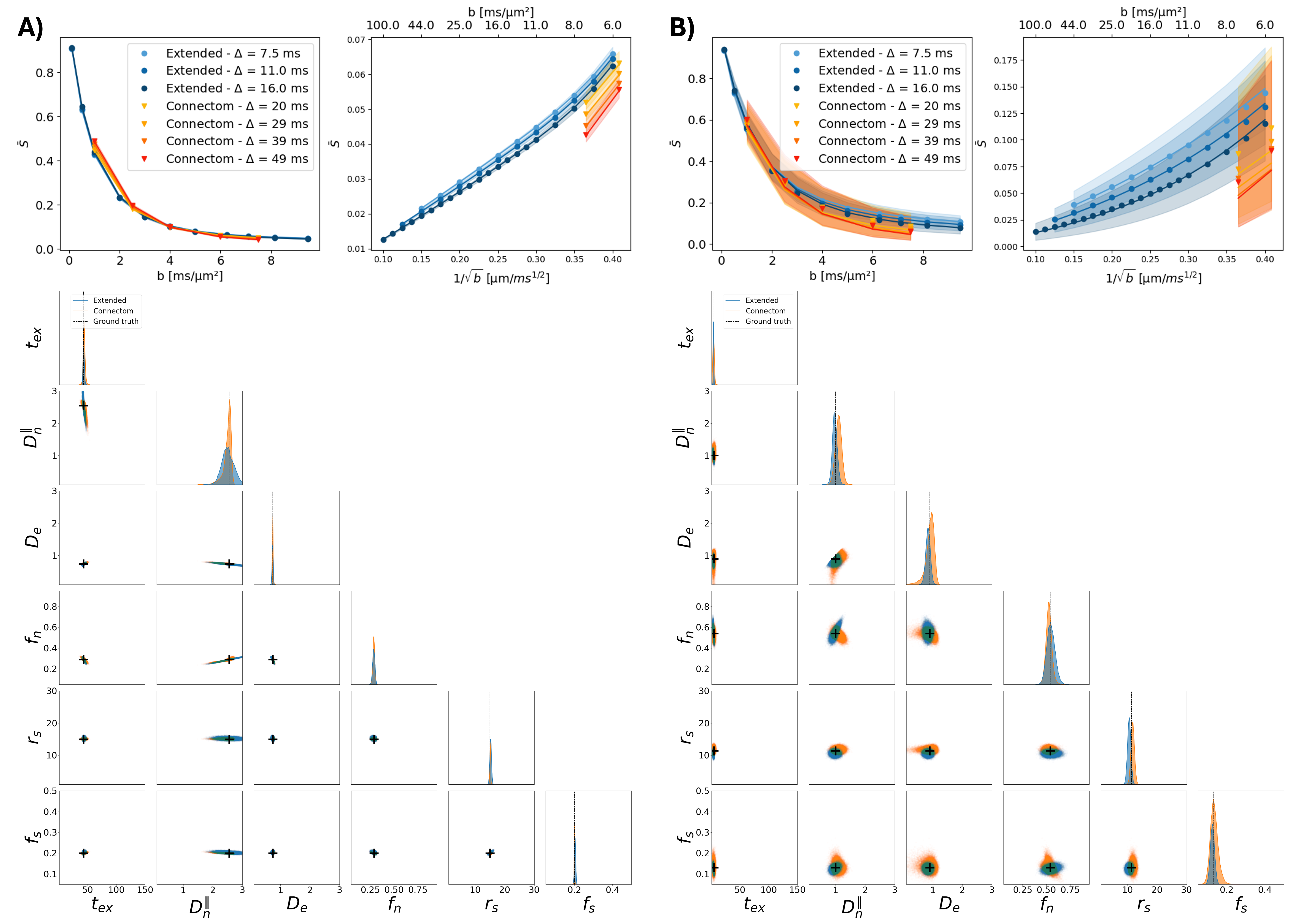 }
\caption{Simulated signals of the SANDIX model, without noise, obtained from exemplar gray matter tissues using both acquisition protocols, and posterior distributions obtained with µGUIDE. For each subplot, upper rows: left plots show the signal with b-values inferior to 10 ms/µm$^2$ as a function of $b$, and right subplot shows the signal for b-values superior to 5 ms/µm$^2$ as a function of $b^{-1/2}$. Dots represent the simulated data, solid lines show signals generated from the MAP estimates, and shaded areas encompass all signals corresponding to parameter combinations sampled from the posterior distributions, illustrating their uncertainty. Bottom rows: Marginal posterior distributions of each parameter obtained with µGUIDE in the diagonal, and pairplots in the lower diagonal. Vertical black dashed lines and black crosses correspond to the ground truth values. Model parameters used to generate the signals are the following: A) $t_{ex}=43$ ms, $D_{n}^{\|}=2.55$ µm${}^2$/ms, $D_e=0.74$ µm${}^2$/ms, $f_n=0.29$, $r_s=15$ µm and $f_s=0.2$ \citep{jelescu_neurite_2022}; B) $t_{ex}=4.95$ ms, $D_{n}^{\|}=1.0$ µm${}^2$/ms, $D_e=0.9$ µm${}^2$/ms, $f_n=0.54$, $r_s=11.4$ µm and $f_s=0.13$ \citep{olesen_diffusion_2022}.}
\label{fig:signals_post_SANDIX}
\end{center}
\end{figure}

Figures~\ref{fig:signals_post_NEXI} and \ref{fig:signals_post_SANDIX} present simulated signals of representative gray matter tissue configurations \citep{jelescu_neurite_2022,olesen_diffusion_2022} under both acquisition protocols, in a noise-free scenario. It is worth noting that the b-values used in the Connectom protocol are not strong enough to reach the stick power law. Additionally, at fixed b-value, the dMRI signal - for the set of model parameters investigated - systematically decreases with diffusion time, reflecting the dominance on time dependence of exchange dynamics over structural restrictions, as enforced in both models \citep{olesen_diffusion_2022}. Posterior distributions obtained using µGUIDE demonstrate accurate and consistent parameter recovery across both protocols and models, with noticeably improved accuracy (i.e., smaller bias in the MAP) and precision (i.e., narrower posterior distributions) when using the extensive \textit{ex vivo} acquisition protocol. The pairplots further illustrate the relationships between parameters, showing weak correlations and compact, unimodal posterior distributions, consistent with good parameter recovery in this noise-free setting.

The difference in uncertainty observed for the NEXI model between scenarios A and B in Figure~\ref{fig:signals_post_NEXI} can be explained by the underlying parameter regimes. In scenario A, the intra- and extra-neurite diffusivities are well separated ($D_{n}^{\|}=2.55$ µm${}^2$/ms and $D_e=0.74$ µm${}^2$/ms), and the exchange time is relatively long ($t_{ex}=43$), leading to more distinct signal contributions from each compartment and facilitating the estimation of $f_n$. In contrast, in scenario B, the diffusivities are closer ($D_{n}^{\|}=1.45$ µm${}^2$/ms and $D_e=0.55$ µm${}^2$/ms), and the exchange time is much shorter ($t_{ex}=8.15$ ms). This results in stronger mixing between compartments with similar diffusion properties, making it more difficult to disentangle their respective contributions and leading to increased uncertainty in $f_n$. Similar trends are observed for the SANDIX model (Figure~\ref{fig:signals_post_SANDIX}). In addition, the longer diffusion times used in the Connectom protocol reduce sensitivity to fast exchange processes, which may further contribute to the increased uncertainty observed in this regime.

\begin{figure}[htbp]
\begin{center}
\includegraphics[width=\textwidth]{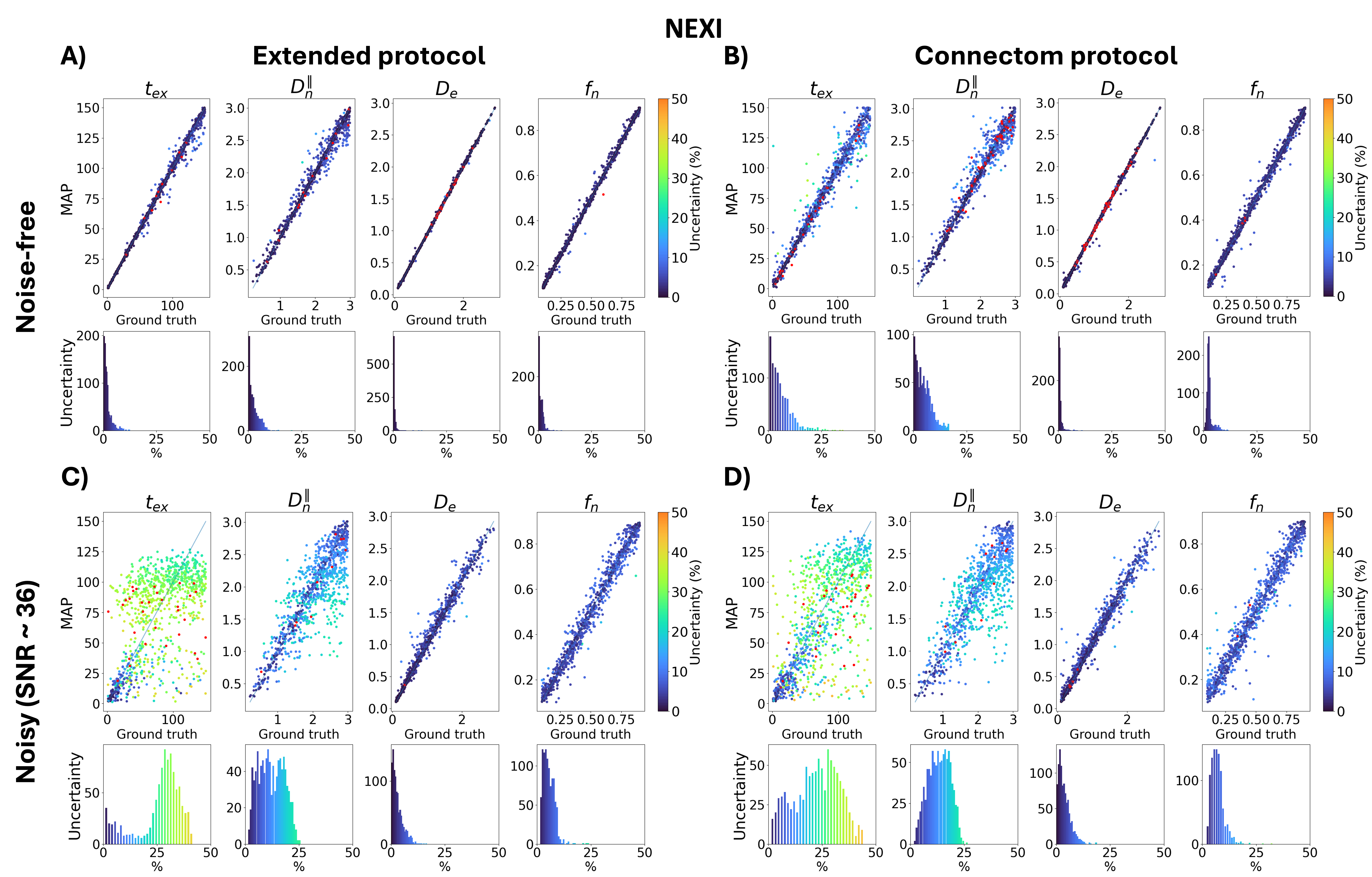}
\caption{Fitting results for the NEXI model using µGUIDE on 1000 test simulations. Results are shown for the extensive \textit{ex vivo} acquisition protocol (A \& C), and the NEXI 3T Connectom protocol (B \& D). In each subplot, the top row displays the MAP estimates of the model parameters plotted against their ground truth values, color-coded by their uncertainty values. Red dots indicate cases where the posterior distribution was identified as degenerate. The bottom row shows the distribution of uncertainty values across all test simulations.}
\label{fig:uGUIDE_fitting_NEXI}
\end{center}
\end{figure}

\begin{figure}[htbp]
\begin{center}
\includegraphics[width=\textwidth]{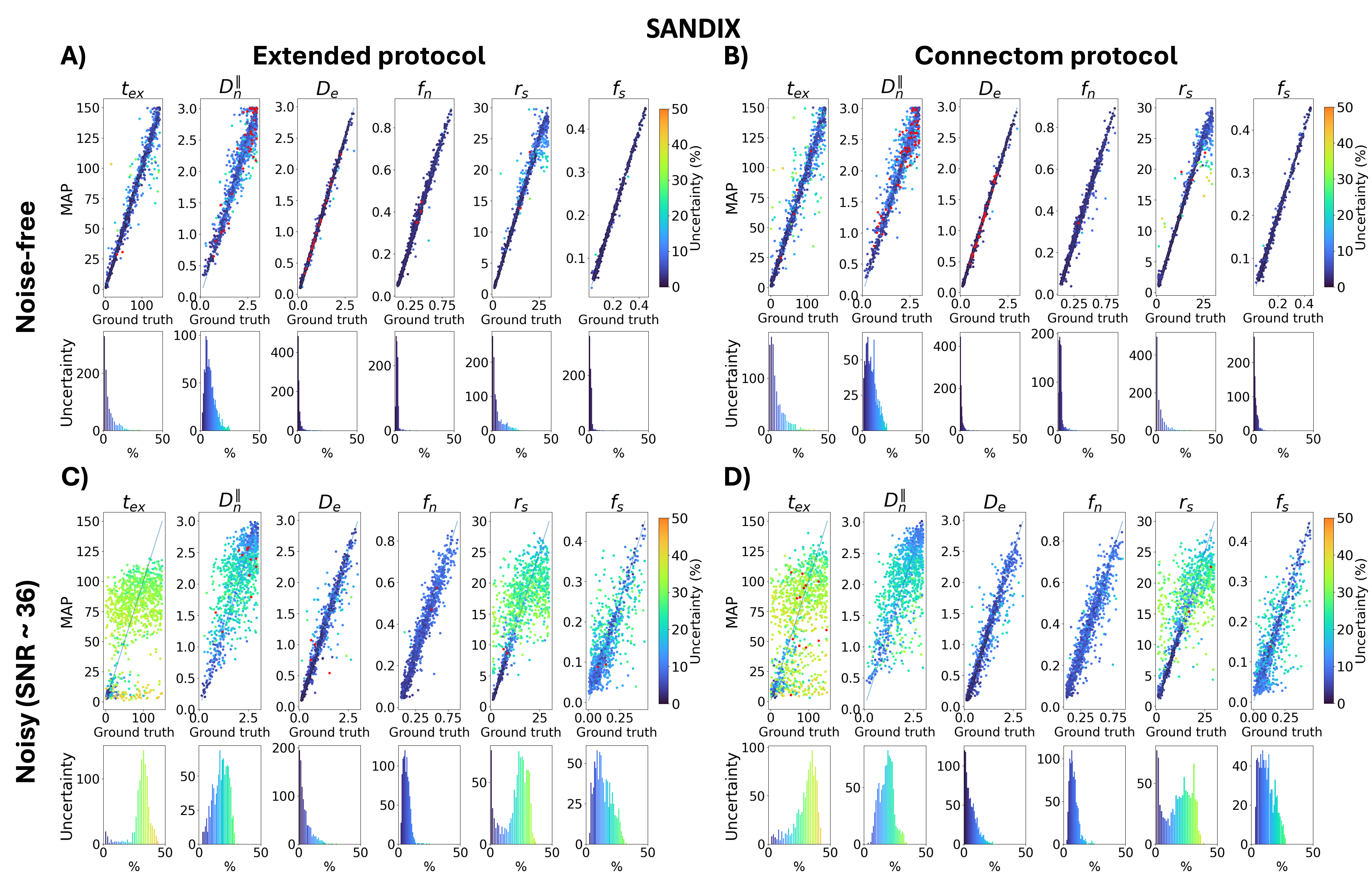}
\caption{Fitting results for the SANDIX model using µGUIDE on 1000 test simulations. Results are shown for the extensive \textit{ex vivo} acquisition protocol (A \& C), and the NEXI 3T Connectom protocol (B \& D). In each subplot, the top row displays the MAP estimates of the model parameters plotted against their ground truth values, color-coded by their uncertainty values. Red dots indicate cases where the posterior distribution was identified as degenerate. The bottom row shows the distribution of uncertainty values across all test simulations.}
\label{fig:uGUIDE_fitting_SANDIX}
\end{center}
\end{figure}

Figures~\ref{fig:uGUIDE_fitting_NEXI} and \ref{fig:uGUIDE_fitting_SANDIX} expand upon the results shown in Figures~\ref{fig:signals_post_NEXI} and \ref{fig:signals_post_SANDIX} by evaluating 1000 test simulations. These figures present µGUIDE fitting results for the NEXI and SANDIX models across both acquisition protocols, under both noise-free (A \& B) and noisy (C \& D) conditions. The extensive \textit{ex vivo} protocol enables highly accurate and precise parameter estimation in the absence of noise, with minimal degeneracies (indicated by red dots) and low uncertainty (subplots A). However, its performance significantly degrades in the presence of noise, as the signal at high b-values is mostly lost (subplots C). In contrast, the Connectom protocol, relying on lower b-values, is less affected by low SNR, but suffers from reduced accuracy and increased variance even in noise-free settings (subplots B and D). Across simulations, we observe that low-uncertainty estimates tend to coincide with low-bias estimates (falling near the diagonal), whereas biased estimates (away from the diagonal) are typically associated with higher uncertainty. This suggests that wider posterior distributions may still encompass the ground truth values, even when MAP estimates are biased. Additionally, the presence of noise appears to reduce the number of detected degeneracies (Table~\ref{tab:degeneracies_simu}). This can be explained by the increased uncertainty caused by noise, which widens the posterior distributions. In this scenario, multiple distinct peaks can blend into a single broader peak, causing degeneracies to be hidden. 

Similar trends are obtained using a non-linear least squares (NLLS) fitting approach with Rician mean correction, as implemented in \citet{uhl_quantifying_2024} (see Supplementary Figures~\ref{fig:suppl_NLLS_fitting_NEXI} and \ref{fig:suppl_NLLS_fitting_SANDIX}). However, some NLLS estimates reach the imposed parameter bounds (e.g., for the NEXI model, $D_n$, $D_e$ at their upper limits and $t_{ex}$, $f_n$ near zero), indicating that the optimizer converged to boundary solutions of the parameter space rather than physiologically meaningful values. Additionally, even in the absence of noise, NLLS estimates for several SANDIX parameters exhibit substantial dispersion, reflecting the presence of multiple parameter combinations yielding similar signals and an optimization landscape likely containing multiple local minima. As a result, the optimizer may converge to different solutions that fit the data equally well. µGUIDE offers several advantages over NLLS, including the detection of degeneracies, the estimation of parameter uncertainty, and computational time between 5 and 12 times faster on simulations (see Supplementary Table~\ref{tab:suppl_time_fitting_uGUIDE_NLLS}).

\begin{table}[h!]
\centering
\caption{Percentage of degenerate voxels for the NEXI and SANDIX models, in noise-free and noisy (SNR$\sim$36) conditions, for both the extensive and Connectom protocols.}
\label{tab:degeneracies_simu}
\setlength{\tabcolsep}{9pt}%
\renewcommand{\arraystretch}{1.2}
\begin{tabular}{| l || c | c | c | c | c | c | c | c |}
\hline
 \ & \ & \ & $t_{ex}$ & $D_{n}^{\|}$ & $D_e$ & $f_n$ & $r_s$ & $f_s$ \\
\hline
  \multirow{4}{*}{NEXI} & \multirow{2}{*}{Extensive protocol} & Noise-free & 1.2 & 1.3 & 2.0 & 0.1 & \ & \ \\
  & & Noisy & 3.1 & 1.0 & 0.1 & 0.0 & \ & \ \\ \cline{2-9}
  & \multirow{2}{*}{Connectom protocol} & Noise-free & 2.5 & 4.2 & 5.1 & 0.5 & \ & \ \\
  & & Noisy & 2.6 & 1.2 & 0.5 & 0.2 & \ & \ \\
  \hline
  \multirow{4}{*}{SANDIX} & \multirow{2}{*}{Extensive protocol} & Noise-free & 0.2 & 3.4 & 2.1 & 0.4 & 0.2 & 0.0 \\
  & & Noisy & 0.0 & 1.2 & 0.4 & 0.2 & 0.2 & 0.2 \\ \cline{2-9}
  & \multirow{2}{*}{Connectom protocol} & Noise-free & 0.4 & 5.6 & 3.2 & 0.1 & 0.3 & 0.0 \\
  & & Noisy & 1.3 & 0.0 & 0.0 & 0.0 & 0.2 & 0.1 \\
\hline
\end{tabular}
\end{table}

As shown in Figure~\ref{fig:uGUIDE_fitting_NEXI}D, estimating the exchange time $t_{ex}$ using the NEXI model under realistic conditions, i.e., using the Connectom acquisition protocol with Rician-distributed noise, is challenging. This is especially true for longer exchange times that exceed the protocol’s range of sampled diffusion times ($\Delta \in [20; 49]$ ms and $\delta=$9 ms), resulting in biased and uncertain estimates. Nonetheless, the NEXI model reliably estimates other key parameters such as $D_e$ and $f_n$.
When applying the SANDIX model to simulations generated using the Connectom protocol and realistic noise levels, Figure~\ref{fig:uGUIDE_fitting_SANDIX}.D demonstrates that the exchange time estimates are mostly unreliable. However, as with NEXI, the model provided robust estimates of $D_e$ and $f_n$. In addition, we note that for soma radii larger than 17 µm, the uncertainty in the estimates increases, which is consistent with previous findings \citep{dong_vivo_2025}.

\subsection{\textit{In vivo} dMRI Data}

\begin{figure}[htbp]
\begin{center}
\includegraphics[width=\textwidth]{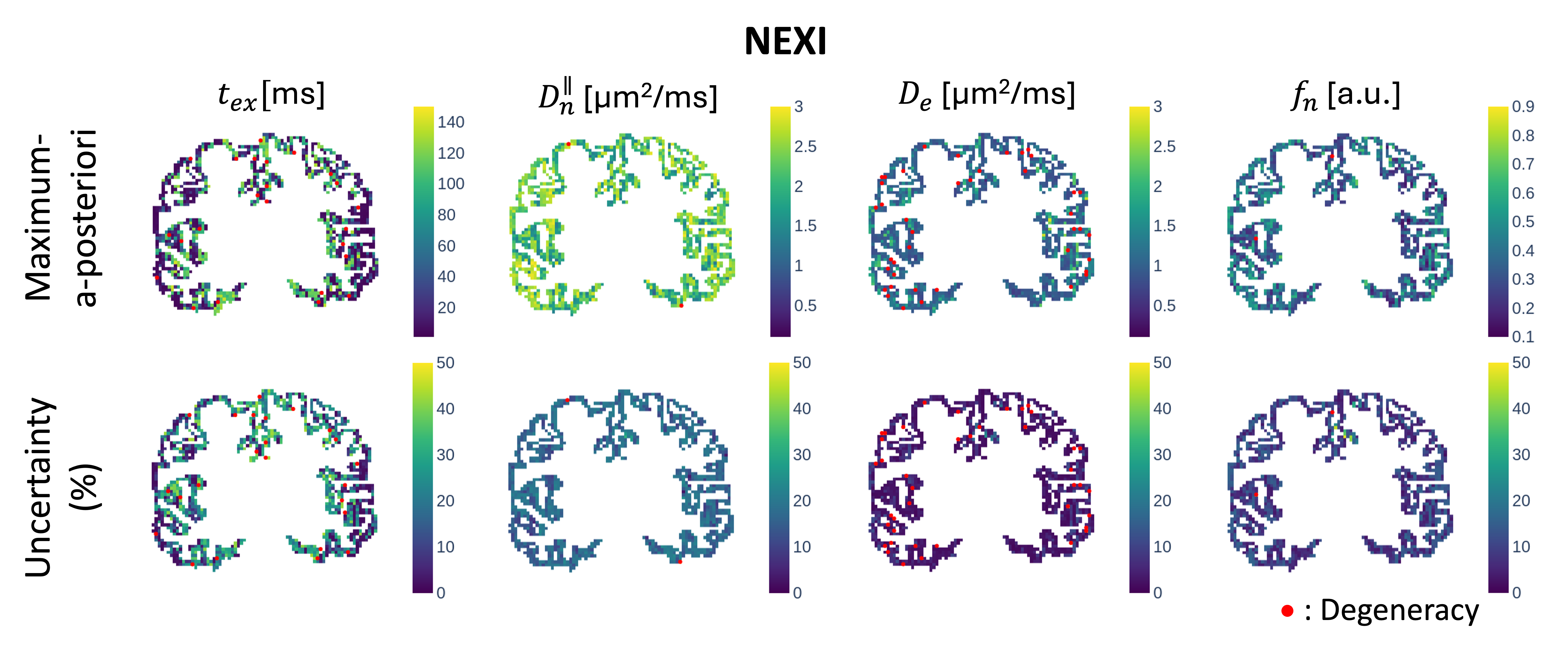}
\caption{Fitting results of the NEXI model on \textit{in vivo} data from one participant scanned using the 3T Connectom protocol, estimated using µGUIDE. For each model parameter, the MAP estimate and associated uncertainty are shown for a single coronal slice of the brain. Voxels exhibiting degenerate posterior distributions are marked with red dots.}
\label{fig:uGUIDE_fitting_participant_NEXI}
\end{center}
\end{figure}

\begin{figure}[htbp]
\begin{center}
\includegraphics[width=\textwidth]{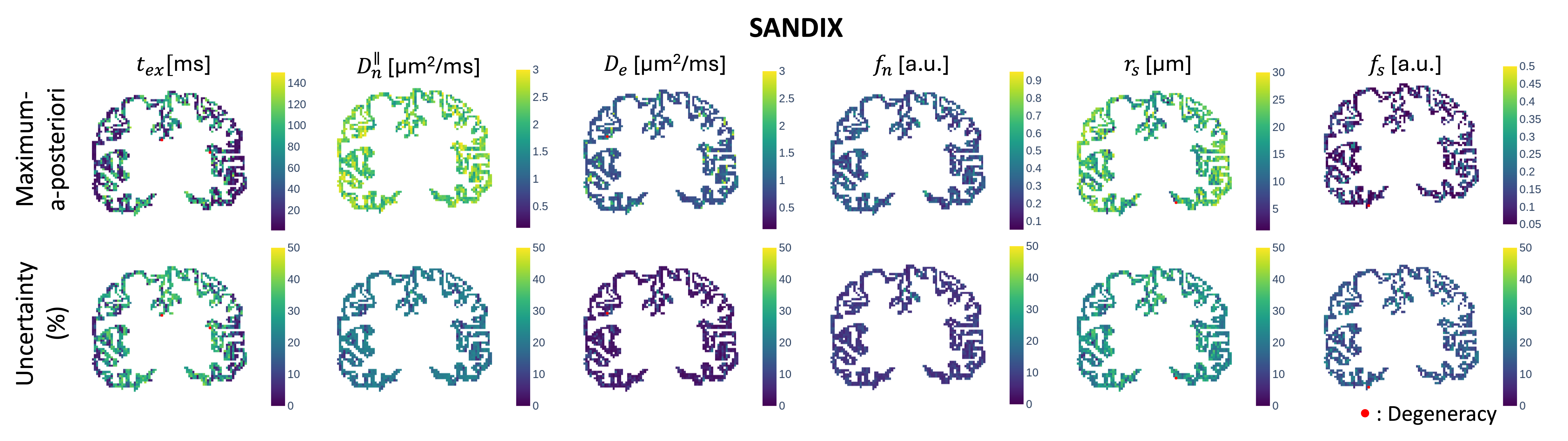}
\caption{Fitting results of the SANDIX model on \textit{in vivo} data from one participant scanned using the 3T Connectom protocol, estimated using µGUIDE. For each model parameter, the MAP estimate and associated uncertainty are shown across a brain slice. Voxels exhibiting degenerate posterior distributions are marked with red dots.}
\label{fig:uGUIDE_fitting_participant_SANDIX}
\end{center}
\end{figure}

Figures~\ref{fig:uGUIDE_fitting_participant_NEXI} and \ref{fig:uGUIDE_fitting_participant_SANDIX} show the parameter estimates obtained using µGUIDE with the NEXI and SANDIX models on \textit{in vivo} data from a participant scanned with the 3T Connectom protocol. Each figure displays both the MAP estimates and associated uncertainty measures for each parameter. Red dots indicate voxels presenting a degeneracy in their posterior distributions. As seen in the simulations, both models yield high uncertainty for the exchange time $t_{ex}$, indicating limited confidence in its estimation with the current protocol. In contrast, estimates of $D_e$ and $f_n$ exhibit low uncertainty, consistent with the model’s robust performance for these parameters in the simulated data.

\begin{figure}[htbp]
\begin{center}
\includegraphics[width=\textwidth]{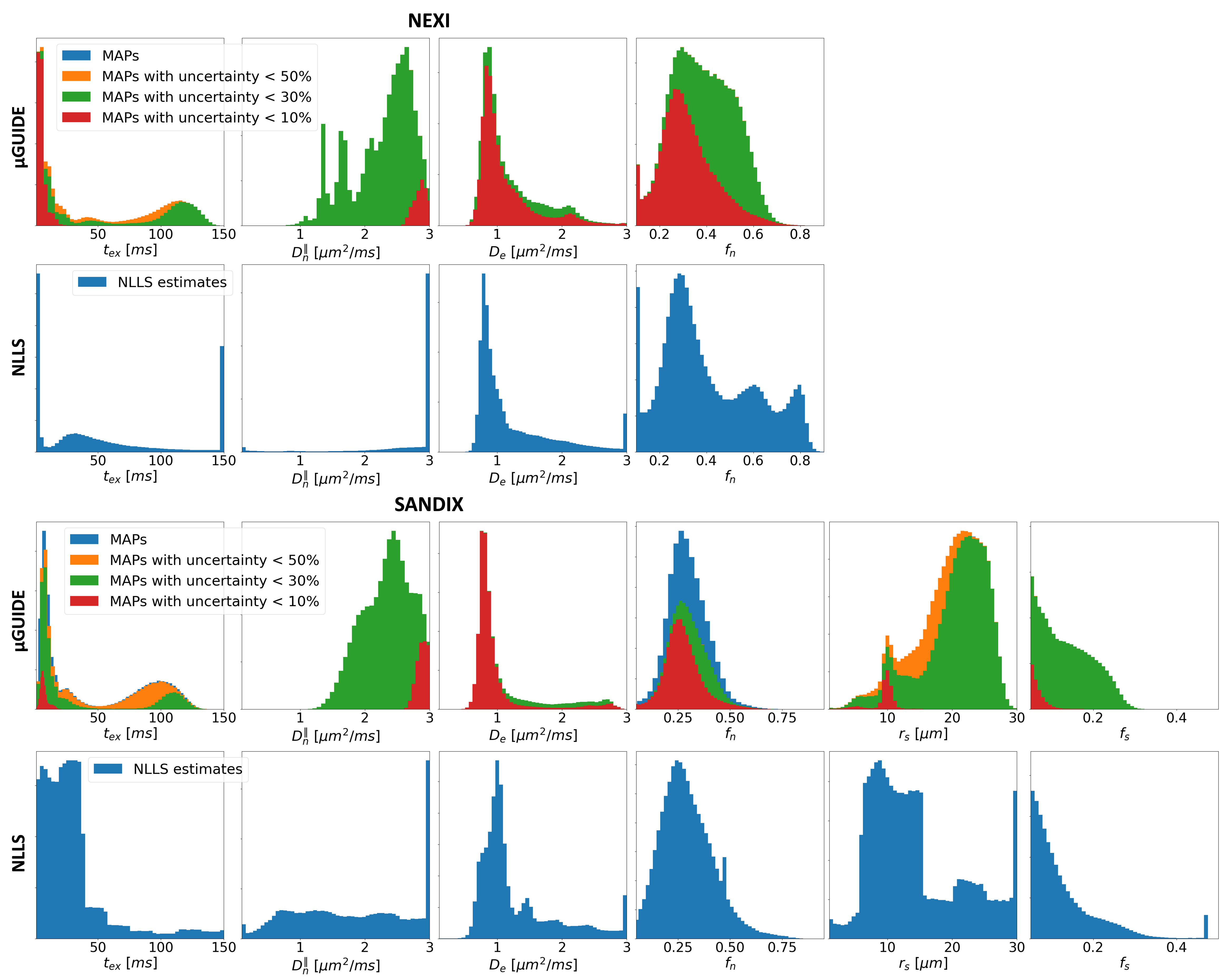}
\caption{Comparison of parameter estimates across cortical ribbons from all participants using µGUIDE and a NLLS method. µGUIDE estimates are thresholded based on posterior distribution uncertainty, with distributions shown for voxels with uncertainty below 50\%, 30\%, and 10\%. In some cases, thresholding does not alter the distribution, resulting in overlapping curves that may not be visually distinguishable.}
\label{fig:uGUIDE_NLLS}
\end{center}
\end{figure}

Figure~\ref{fig:uGUIDE_NLLS} presents a comparison of NEXI and SANDIX parameter estimates across the cortical ribbons of all participants, obtained using µGUIDE and the NLLS method, as implemented in \citet{uhl_quantifying_2024}. For µGUIDE, estimates are thresholded by their posterior uncertainty: we report distributions retaining only the voxels whose uncertainty is inferior to 50\%, 30\%, and 10\%. Voxels with the lowest uncertainty are considered the most reliable. In contrast, no quality-based filtering is applied to the NLLS results. Notably, NLLS estimates frequently hit the boundaries of the predefined parameter ranges, indicating potential instability.

Table~\ref{tab:degeneracies} reports the proportion of voxels exhibiting posterior distribution degeneracies across all six acquisitions for both models. Table~\ref{tab:uncertainties} summarizes the percentage of voxels with uncertainty below 50\%, 30\%, and 10\%, respectively, again for both models.

\begin{table}[h!]
\centering
\caption{Percentage of degenerate voxels across cortical ribbons from all data acquisitions.
\label{tab:degeneracies}}
\setlength{\tabcolsep}{20pt}%
\renewcommand{\arraystretch}{1.4}
\begin{tabular}{| l || c | c | c | c | c | c |}
\hline
 \  & $t_{ex}$ & $D_{n}^{\|}$ & $D_e$ & $f_n$ & $r_s$ & $f_s$ \\
\hline
  NEXI  & 1.76 & 0.52 & 4.23 & 0.11 & \ & \ \\
  SANDIX  & 0.47 & 0.02 & 0.27 & 0.01 & 0.14 & 0.25 \\
\hline
\end{tabular}
\end{table}

\begin{table}[h!]
\centering
\caption{Percentage of voxels across cortical ribbons from all data acquisitions whose uncertainty is inferior to 50\%, 30\%, 20\% and 10\%.
\label{tab:uncertainties}}
\setlength{\tabcolsep}{9pt}%
\renewcommand{\arraystretch}{1.2}
\begin{tabular}{| l || c | c | c | c | c | c | c |}
\hline
 \  & \ & $t_{ex}$ & $D_{n}^{\|}$ & $D_e$ & $f_n$ & $r_s$ & $f_s$ \\
\hline
  \multirow{2}{*}{NEXI}  & Uncertainty $<$ 50\% & 99.98 & 100 & 100 & 100 & \ & \ \\
  & Uncertainty $<$ 30\% & 82.37 & 99.95 & 100 & 99.78 & \ & \ \\
  & Uncertainty $<$ 20\% & 51.18 & 85.44 & 99.87 & 99.29 & \ & \ \\
  & Uncertainty $<$ 10 & 40.85 & 8.71 & 91.72 & 51.23 & \ & \ \\
  \hline
  \multirow{2}{*}{SANDIX}  & Uncertainty $<$ 50\% & 100 & 100 & 100 & 100 & 100 & 100 \\
  & Uncertainty $<$ 30\% & 62.90 & 99.98 & 99.99 & 100 & 83.12 & 99.96 \\
  & Uncertainty $<$ 20\% & 38.5 & 52.12 & 99.15 & 99.79 & 31.51 & 82.02 \\
  & Uncertainty $<$ 10\% & 26.73 & 11.80 & 85.36 & 69.61 & 3.45 & 15.39 \\
\hline
\end{tabular}
\end{table}

Estimates of the exchange time $t_{ex}$ display a bimodal distribution, with peaks at both low and high values (Figure~\ref{fig:uGUIDE_NLLS}). The high $t_{ex}$ peak may reflect partial volume effects from WM, where myelin reduces the axons permeability \citep{dong_vivo_2025}. As expected from simulations, this peak disappears when filtering out voxels with high uncertainty. Only 40.85\% (NEXI) and 26.73\% (SANDIX) of voxels yield reliable $t_{ex}$ estimates with uncertainty below 10\% (corresponding to an interquartile range of $\sim$15 ms) (Table~\ref{tab:uncertainties}). With this uncertainty threshold, the mean MAP estimates are 5.51 ms (NEXI) and 7.24 ms (SANDIX), which reflect rapid water exchange across permeable neurite membranes.

For intra-neurite diffusivity $D_{n}^{\|}$, µGUIDE and NLLS produce noticeably different distributions (Figure~\ref{fig:uGUIDE_NLLS}). Only $\sim$10\% of the $D_{n}^{\|}$ estimates are deemed reliable with µGUIDE, with a mean MAP of 2.85 µm${}^2$/ms for NEXI and 2.88 µm${}^2$/ms for SANDIX (Table~\ref{tab:uncertainties}).

By contrast, estimates of extra-neurite diffusivity $D_e$ and neurite signal fraction $f_n$ show more consistency for most voxels across both methods and models, although $D_e$ distributions with NLLS present longer tail for SANDIX, and higher $f_n$ estimates for NEXI (Figure~\ref{fig:uGUIDE_NLLS}). These parameters also exhibit the lowest uncertainty in the µGUIDE estimates, in agreement with simulation findings (Figures~\ref{fig:uGUIDE_fitting_NEXI} \& \ref{fig:uGUIDE_fitting_SANDIX}).

Soma radii estimates ($r_s$) show substantial differences between µGUIDE and NLLS. As predicted by simulations, estimates of large soma radii tend to have higher uncertainty (Figure~\ref{fig:uGUIDE_fitting_SANDIX}), causing the high-radius peak to disappear when filtering for voxels with less than 10\% uncertainty (equivalent to a $\sim$3 µm interquartile range). Only 3.45\% of voxels yield reliable $r_s$ estimates at this threshold (Table~\ref{tab:uncertainties}), clustering around one main peak at $\sim$10 µm, and a smaller one $\sim$5 µm. Estimates of the soma signal fraction $f_s$ are also uncertain, with only 15.39\% of voxels passing the 10\% uncertainty threshold. For those few voxels, the mean estimate of $f_s$ is 0.05.

Scan–rescan analyses for two participants show that the proportions of degenerate voxels (Table~\ref{tab:degeneracies_scan_rescan}) and the percentages of voxels with uncertainty below 50\%, 30\%, and 10\% (Table~\ref{tab:uncertainties_scan_rescan}) are highly consistent across sessions for both models. This reproducibility suggests that the degeneracy detection and uncertainty quantification provided by µGUIDE are robust and reproducible.

\begin{table}[h!]
\centering
\caption{Percentage of degenerate voxels across cortical ribbons from the scan-rescan acquisitions for both subjects (S1 and S2).
\label{tab:degeneracies_scan_rescan}}
\setlength{\tabcolsep}{8pt}%
\renewcommand{\arraystretch}{1}
\begin{tabular}{| l || c | c | c | c | c | c | c | c | c | c | c | c |}
\hline
 \  & \multicolumn{2}{c|}{$t_{ex}$} & \multicolumn{2}{c|}{$D_{n}^{\|}$} & \multicolumn{2}{c|}{$D_e$} & \multicolumn{2}{c|}{$f_n$} & \multicolumn{2}{c|} {$r_s$} & \multicolumn{2}{c|}{$f_s$} \\
 \hline
 \ & S1 & S2 & S1 & S2 & S1 & S2 & S1 & S2 & S1 & S2 & S1 & S2 \\
 \hline
 NEXI  & 1.77 & 1.77 & 0.42 & 0.54 & 4.20 & 4.51 & 0.09 & 0.13 & \ & \ & \ & \ \\
 SANDIX  & 0.42 & 0.52 & 0.03 & 0.02 & 0.23 & 0.26 & 0.01 & 0.01 & 0.14 & 0.13 & 0.20 & 0.21 \\
\hline
\end{tabular}
\end{table}

\begin{table}[h!]
\centering
\caption{Percentage of voxels across cortical ribbons from the scan-rescan acquisitions for both subjects (S1 and S2) whose uncertainty is inferior to 50\%, 30\% and 10\%.
\label{tab:uncertainties_scan_rescan}}
\begin{adjustbox}{width=\textwidth}
\begin{tabular}{| l || c | c | c | c | c | c | c | c | c | c | c | c | c |}
  \hline
  \multicolumn{2}{|c|}{} & \multicolumn{2}{c|}{$t_{ex}$} & \multicolumn{2}{c|}{$D_{n}^{\|}$} & \multicolumn{2}{c|}{$D_e$} & \multicolumn{2}{c|}{$f_n$} & \multicolumn{2}{c|} {$r_s$} & \multicolumn{2}{c|}{$f_s$} \\
  \hline
  \multicolumn{2}{|c|}{} & S1 & S2 & S1 & S2 & S1 & S2 & S1 & S2 & S1 & S2 & S1 & S2 \\
  \hline
  \multirow{2}{*}{NEXI} & Uncertainty $<$ 50\% & 99.99 & 99.99 & 100 & 100 & 100 & 100 & 100 & 100 & \ & \ & \ & \ \\
  & Uncertainty $<$ 30\% & 82.32 & 82.49 & 99.96 & 99.95 & 99.99 & 100 & 99.82 & 99.79 & \ & \ & \ & \ \\
  & Uncertainty $<$ 10\% & 42.16 & 40.58 & 7.95 & 9.22 & 91.77 & 91.39 & 49.78 & 51.51 & \ & \ & \ & \ \\
  \hline
  \multirow{2}{*}{SANDIX}  & Uncertainty $<$ 50\% & 100 & 100 & 100 & 100 & 100 & 100 & 100 & 100 & 100 & 100 & 100 & 100 \\
  & Uncertainty $<$ 30\% & 64.18 & 62.92 & 99.98 & 99.98 & 99.99 & 99.99 & 99.99 & 100 & 83.39 & 83.27 & 99.97 & 99.96 \\
  & Uncertainty $<$ 10\% & 27.87 & 27.15 & 11.22 & 11.57 & 85.36 & 85.81 & 68.42 & 69.82 & 3.21 & 3.29 & 14.41 & 16.08 \\
\hline
\end{tabular}
\end{adjustbox}
\end{table}

\begin{figure}[htbp]
\begin{center}
\includegraphics[width=\textwidth]{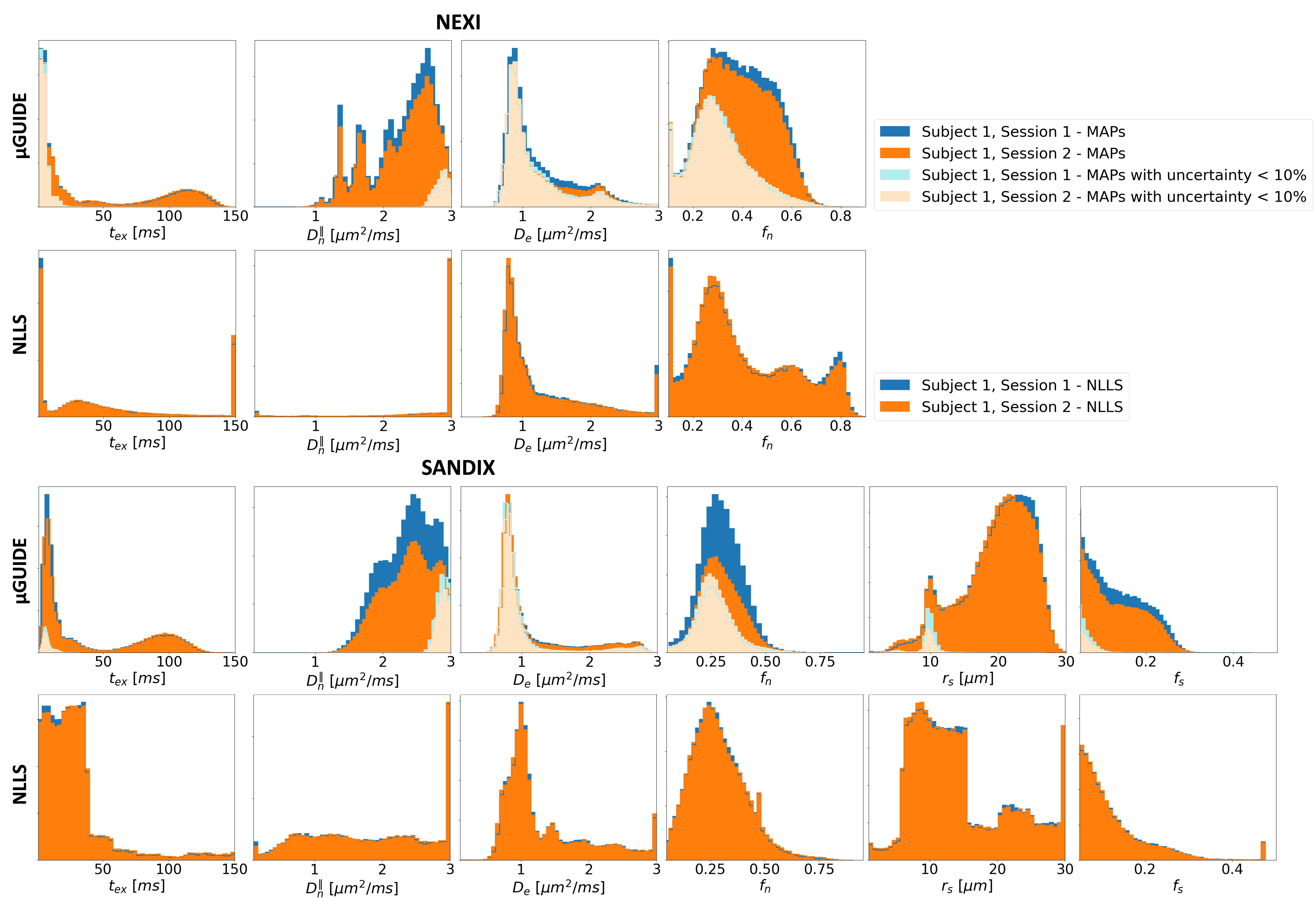}
\caption{Comparison of parameter estimates across cortical ribbon from one participant on two sessions using µGUIDE and a NLLS method. µGUIDE estimates are thresholded based on posterior distribution uncertainty, with distributions shown for voxels with uncertainty below 10\%.}
\label{fig:uGUIDE_NLLS_scan_rescan}
\end{center}
\end{figure}

Figure~\ref{fig:uGUIDE_NLLS_scan_rescan} replicates the comparison from Figure~\ref{fig:uGUIDE_NLLS} using scan–rescan data from two participants (see Figure~\ref{fig:suppl_uGUIDE_NLLS_scan_rescan} for results on the second participant scanned twice). For µGUIDE, we show both the full distribution of MAP estimates and the subset of MAPs with uncertainty below 10\%. Overall, both µGUIDE and NLLS yield similar parameter distributions and appear robust across sessions. The small differences observed between scan and rescan are likely attributable to noise effects. In particular, the largest variability is observed for $D_{n}^{\|}$, which is consistent with this parameter showing higher uncertainty in both models.

\section{Discussion}
In this work, we investigated the reliability of parameter estimates in GM biophysical models. Specifically, we examined the NEXI and SANDIX models, which have been proposed to account for exchange between neurites and the extra-cellular space. While these models hold promise for advancing our understanding of GM microstructure, their solution landscapes and the reliability of their parameter estimates have not been thoroughly assessed.
Our results reveal that certain parameters, including exchange time, intra-neurite diffusivity and soma radius and fraction, are challenging to estimate accurately. We show that the estimated values can be unreliable, either because multiple parameter combinations can explain equally well the observed data (i.e., degeneracies), or because the confidence in the estimates is inherently low (i.e., high uncertainty). These issues have significant implications for the interpretation of results based on these models. Thus, this study emphasizes the importance of quantifying the reliability of model parameter estimates, using Bayesian inference. We suggest highlighting degeneracies and uncertainties, e.g. as provided by µGUIDE, to complement measures of MAP to refine and improve the interpretability of the biophysical model parameter estimates, ultimately leading to a more reliable understanding of the biological changes occurring in pathology and disease.

\subsection{Choice of Acquisition Protocol and Impact of Noise}
Our results demonstrate that extensively sampling the acquisition space across a broad range of b-values and diffusion times substantially reduces parameter degeneracies and leads to lower uncertainties. Simulations (Figures~\ref{fig:uGUIDE_fitting_NEXI}\&\ref{fig:uGUIDE_fitting_SANDIX}) show that, in a noise-free scenario, the extensive protocol results in a low proportion of degenerate posterior distributions (1.2\% for NEXI and 0.2\% for SANDIX for $t_{ex}$) and yields estimates with low uncertainty (inferior to 20\% for NEXI and 42\% for SANDIX, for all parameters). In contrast, the reduced Connectom protocol, while still yielding reasonably accurate estimates, exhibits a higher incidence of degeneracies and greater uncertainty. Specifically, degeneracies are 2 times more likely to occur using the Connectom protocol compared to the extensive one for the exchange time with both models, and uncertainties are almost three times higher for $t_{ex}$ with the NEXI model. Interestingly, we also observe that NEXI exhibits a higher proportion of degeneracies than SANDIX, both in simulations and in \textit{in vivo} data (Tables~\ref{tab:degeneracies_simu}\&\ref{tab:degeneracies}). Importantly, this difference should not be interpreted as evidence that SANDIX is intrinsically more accurate. Rather, degeneracies reflect the ability of different parameter combinations to produce similar signals given both the model structure and the acquisition protocol. The simpler two-compartment structure of NEXI may lead to stronger parameter coupling given those acquisition protocols, where different combinations of exchange time and diffusivities produce similar signals, resulting in increased degeneracy. In contrast, the additional compartment in SANDIX can help capture specific signal features and partially reduce degeneracies in certain regimes. However, this increased model complexity does not fully resolve parameter degeneracies and instead leads to high uncertainty frequent degeneracies in the additional parameters related to the soma. This suggests that, in practice, the added flexibility of SANDIX does not necessarily translate into more reliable microstructural information in this setting. Instead, it emphasizes the importance of matching model complexity to the sensitivity of the acquisition protocol. Under limited acquisition conditions, simpler models such as NEXI may provide more robust and interpretable estimates, whereas more complex models require richer sampling strategies to ensure reliable parameter estimates.

Our results highlight the value of richer sampling for improving the robustness and reliability of parameter estimation in biophysical modeling. However, translating such rich sampling schemes to human \textit{in vivo} studies remains challenging. The Connectom protocol already approaches the upper limit of acceptable scan duration for human participants (45 minutes), and the preclinical protocol is far too long (100h and 14 minutes) to be feasible in this context. In clinical settings, where substantially shorter acquisition times are required, this challenge becomes even more pronounced. Reducing acquisition time would likely further degrade parameter estimation, increasing both uncertainty and degeneracy, particularly for parameters that already exhibit high uncertainty (e.g., exchange time and soma-related parameters). In addition to scan time constraints, hardware limitations, and particularly the achievable gradient strengths, restrict the accessible range of b-values and diffusion times, further limiting sensitivity to some model parameters. Importantly, shortening a protocol is not a trivial subsampling problem, as it requires re-optimizing the acquisition to preserve sensitivity to the parameters of interest. This underscores the need for dedicated protocol optimization strategies tailored to clinical constraints (e.g. \citet{alexander_general_2008,planchuelo-gomez_optimisation_2024,uhl_reduced_2025,popolo_subset_2025}), potentially guided by uncertainty and degeneracy measures \citep{popolo_subset_2025}.

We further show that the presence of realistic (Rician-distributed) noise in the observed signals increases both the bias and uncertainty of parameter estimates, which in turn impact results interpretation. Estimates of large exchange times are particularly impacted, suggesting a reduced sensitivity to intermediate-to-long exchange times beyond the range of sampled diffusion times of the acquisition protocol. This reflects the “filtering” effect of the diffusion-time range, where only exchange processes with characteristic times comparable to the sampled diffusion times can be reliably estimated \citep{jelescu_neurite_2022,dong_vivo_2025}. When diffusion times are too short, slow exchange cannot be captured, causing exchange-time estimates to plateau once the ground truth exceeds the accessible diffusion-time window. Nevertheless, if biologically plausible exchange times in gray matter are on the order of a few tens of milliseconds, reduced sensitivity to much longer values may be less problematic in practice. While Rician noise was used to mimic realistic magnitude dMRI acquisitions, additional simulations using Gaussian noise with matched SNR (see Supplementary Figure~\ref{fig:suppl_uGUIDE_NEXI_Gaussian}) yielded very similar uncertainty and degeneracy patterns, with only slightly reduced bias in the estimated parameters, as expected given the absence of the Rician noise floor at low signal levels. This indicates that the main effects observed here are primarily driven by the signal-to-noise ratio and acquisition design rather than the specific noise distribution.

Overall, these results highlight the critical influence of acquisition design, particularly b-value and diffusion time selection, as well as noise levels, on the accuracy and precision of biophysical parameter estimates. They underscore the need for designing acquisition protocols specifically optimized to enhance the reliability of parameter estimation in gray matter modeling. Future work should focus on refining these protocols to balance scan time, signal quality, and parameter sensitivity, using optimization methods such as \citet{alexander_general_2008,planchuelo-gomez_optimisation_2024,uhl_reduced_2025,popolo_subset_2025}. Incorporating uncertainty and degeneracy measures into this optimization process could further guide the selection of acquisition parameters to maximize the robustness of parameter estimates.

\subsection{Importance of Characterizing Degeneracy and Uncertainty}
Accurate interpretation of model-derived parameters is essential, as misinterpretations can lead to false conclusions. In this context, degeneracies and uncertainties as obtained from posterior distributions are useful complementary information that guide the interpretation of model estimates. Voxels with low posterior uncertainty indicate higher confidence in the corresponding parameter estimates, enabling the identification of robust trends and a more reliable understanding of the underlying tissue microstructure.

Although µGUIDE and NLLS fitting yield similar distributions for certain parameters such as $D_e$ and $f_n$, notable differences emerge for other parameters (Figure~\ref{fig:uGUIDE_NLLS}). In particular, µGUIDE produces lower MAP estimates of exchange time than previously reported using NLLS in \textit{in vivo} studies: 5.51 ms with NEXI and 7.24 ms with SANDIX (mean MAP estimates of voxels with non-degenerate posterior distributions with uncertainty inferior to 10\%), compared to estimates between 10 and 50 ms typically reported in the literature \citep{jelescu_neurite_2022,dong_vivo_2025,uhl_quantifying_2024,veraart_noninvasive_2020}. The high peak observed around $\sim$125 ms in µGUIDE results may be driven by partial volume effects with white matter (and/or CSF), as previously suggested \citep{dong_vivo_2025}.

The exchange-time estimates obtained in this study depend strongly on the uncertainty threshold applied. While estimates obtained with relaxed thresholds are consistent with previously reported values in the human cortex ($\approx$10–80 ms), more conservative thresholds lead to substantially lower mean values. For example, for the NEXI model, the mean MAP decreases from 46.26 ms ($<$50\%) to 43.25 ms ($<$30\%), 24.20 ms ($<$20\%), and 5.51 ms ($<$10\%). A similar trend is observed for SANDIX, with mean values decreasing from 45.58 ms ($<$50\%) to 33.39 ms ($<$30\%), 8.80 ms ($<$20\%), and 7.24 ms ($<$10\%). This reflects a selection effect, whereby only voxels with deemed reliable estimates are retained as the threshold becomes smaller. This filtering excludes estimates with longer exchange times, which are more difficult to estimate reliably, thereby shifting the mean toward lower values. This effect is also closely related to the diffusion-time range of the acquisition protocol, which acts as a filter on the exchange times that can be reliably probed. In particular, the diffusion times used in this study ($\sim$20–50 ms) limit sensitivity to longer exchange processes, resulting in higher uncertainty for large exchange times. Consequently, these values are more likely to be excluded by uncertainty-based filtering, leading to an apparent bias toward shorter exchange times. In comparison, NLLS estimates (54.07 ms for NEXI and 30.20 ms for SANDIX) fall within a similar range but do not account for uncertainty. These results highlight the importance of jointly considering acquisition design and uncertainty when interpreting exchange-time estimates. By leveraging degeneracy detection and uncertainty quantification from µGUIDE, unreliable voxels can be filtered out, isolating only the most trustworthy estimates. Consequently, the parameter distributions obtained via µGUIDE are more robust and interpretable than those derived from NLLS alone, reinforcing the importance of accounting for uncertainty and degeneracy in biophysical modeling.

\subsection{Robustness and Reproducibility}

Our scan–rescan analysis (Figure \ref{fig:uGUIDE_NLLS_scan_rescan}) provides further insight into the robustness of parameter estimation using both µGUIDE and NLLS. Overall, the distributions of estimates were consistent across the two sessions for both methods, indicating good reproducibility. For µGUIDE, the agreement between sessions improved when restricting the analysis to voxels with low posterior uncertainty ($<$ 10\%), highlighting the utility of uncertainty-based filtering to improve reliability. The differences observed between scan and rescan can largely be attributed to noise variability, with the intracellular diffusivity ($D_{n}^{\|}$) showing the largest fluctuations, which is consistent with its higher uncertainty in both NEXI and SANDIX. These results suggest that while both µGUIDE and NLLS yield stable estimates at the subject level, incorporating uncertainty measures, as in µGUIDE, can improve the robustness of reproducibility assessments and help identify parameters more sensitive to noise.

\subsection{Limitations}
The current study has several limitations. First, we evaluated only two acquisition protocols: an extensive protocol developed for \textit{ex vivo} acquisitions, and a protocol feasible on Connectom scanners. Neither reflects the typical scan parameters used in most clinical 3T MRI systems, which are constrained by lower gradient amplitudes and longer echo times, resulting in reduced achievable b-values and lower SNR per unit time \citep{uhl_human_2025}. To ensure a fair comparison between protocols, we added Rician noise at an identical level to both sets of simulated signals, even though this SNR level is unrealistically low for \textit{ex vivo} acquisitions. This choice was made to isolate and study the impact of noise on model fitting performance, despite the fact that the extensive protocol is impractical \textit{in vivo} due to its duration.
Additionally, partial volume effects could have affected parameter estimates in the GM ribbon due to the limited resolution of the acquired images. Future studies could mitigate these effects by leveraging emerging acquisition techniques, such as the DW-GRASE sequence with 3D navigator \citep{li_diffusion-weighted_2025}, which offer higher resolution for time-dependent dMRI.

Second, we focused exclusively on two biophysical models that incorporate exchange times, i.e. NEXI and SANDIX. Other biophysical models also include exchange time, such as SMEX (see Section~\ref{subsec:SANDIX}), or eSANDIX \citep{olesen_diffusion_2022}, which also integrates impermeable neurites. In this study, our goal was not to determine which model best describes gray matter microstructure, but rather to demonstrate that parameter estimates from such models can be prone to degeneracies and uncertainty. This highlights a key limitation in current fitting approaches, where confidence in parameter estimates is not quantified and accounted for when interpreting the results, and degeneracies are ignored \citep{jallais_introducing_2024}.

Third, this study focused exclusively on PGSE acquisitions and did not consider alternative diffusion encoding strategies that can help alleviate parameter degeneracies. Advanced diffusion waveforms can provide increased sensitivity to specific microstructural features and help disentangle the effects of restriction and exchange \citep{chakwizira_diffusion_2023,chakwizira_diffusion_2023-1,chakwizira_diffusion_2025,lasic_tuned_2024}. Moreover, combining multiple complementary acquisition schemes, such as single and double diffusion encoding as proposed in \citet{jallais_combining_2025}, can further improve parameter estimation, enhance the robustness of the fitting, and reduce degeneracies. Future work should explore these strategies to effectively mitigate degeneracies.

\section{Conclusion}
In this study, we investigated the reliability of parameter estimates from biophysical models of gray matter that incorporate water exchange, with a focus on NEXI and SANDIX. 
Through simulations and \textit{in vivo} data, we demonstrated that key parameters, such as exchange time, soma radius and soma fraction, are challenging to estimate reliably, due to intrinsic model degeneracies, sensitivity to acquisition settings, and noise. While extensive protocols can substantially reduce uncertainty and the presence of degeneracies, they remain impractical for \textit{in vivo} acquisitions. In contrast, protocols feasible for \textit{in vivo} human applications, such as the NEXI 3T Connectom protocol, yield more robust fits in the presence of noise, but show reduced accuracy and increased parameter uncertainty.
We showed that µGUIDE provides more robust and interpretable estimates than traditional NLLS fitting by identifying degenerate solutions and leveraging uncertainty measures derived from the posterior distributions. These findings highlight the importance of accounting for the reliability of individual estimates in biophysical modeling, particularly for drawing meaningful biological conclusions and comparing clinical populations.
Ultimately, our work advocates for uncertainty-aware inference as a critical step toward more reliable and biologically meaningful diffusion MRI analysis.

\section*{Data and Code Availability}

The code used in this study is available at: \href{https://github.com/mjallais/NEXI_SANDIX_uGUIDE}{https://github.com/mjallais/NEXI\_SANDIX\_uGUIDE}.
The data used here are available upon request after signing a formal data sharing agreement and providing approval from the requesting researcher's local ethics committee.

\section*{Author Contributions}

Conceptualization: M.J., Q.U., I.J., M.P.;
Data curation: M.J., Q.U., T.P., M.M.;
Formal analysis: M.J., Q.U.;
Funding acquisition: D.K.J., I.J., M.P.;
Investigation: M.M., I.J., M.P.;
Methodology: M.J., Q.U., I.J., M.P.;
Supervision: I.J., M.P.;
Visualization: M.J., Q.U.;
Writing-original draft: M.J.;
Writing-review \& editing: M.J., Q.U., M.M., D.K.J., I.J., M.P.;

\section*{Funding}

MJ and MP are supported by UKRI Future Leaders Fellowship (MR/T020296/2 and 1073). QU, TP and IJ are supported by an SNSF Eccellenza Fellowship (194260). This research was funded in whole, or in part, by  a Wellcome Trust Strategic Award (104943/Z/14/Z). DKJ is also supported by Wellcome Discovery Awards (227882/Z/23/Z and 317797/Z/24/Z). For the purpose of open access, the author has applied a CC BY public copyright licence to any Author Accepted Manuscript version arising from this submission.

\section*{Declaration of Competing Interests}

The authors declare no competing interest.

\newpage
\section*{Supplementary Material}
\setcounter{figure}{0}
\setcounter{table}{0}

\subsection*{NLLS Fitting Performance}

\begin{figure}[!htbp]
\begin{center}
\includegraphics[width=\textwidth]{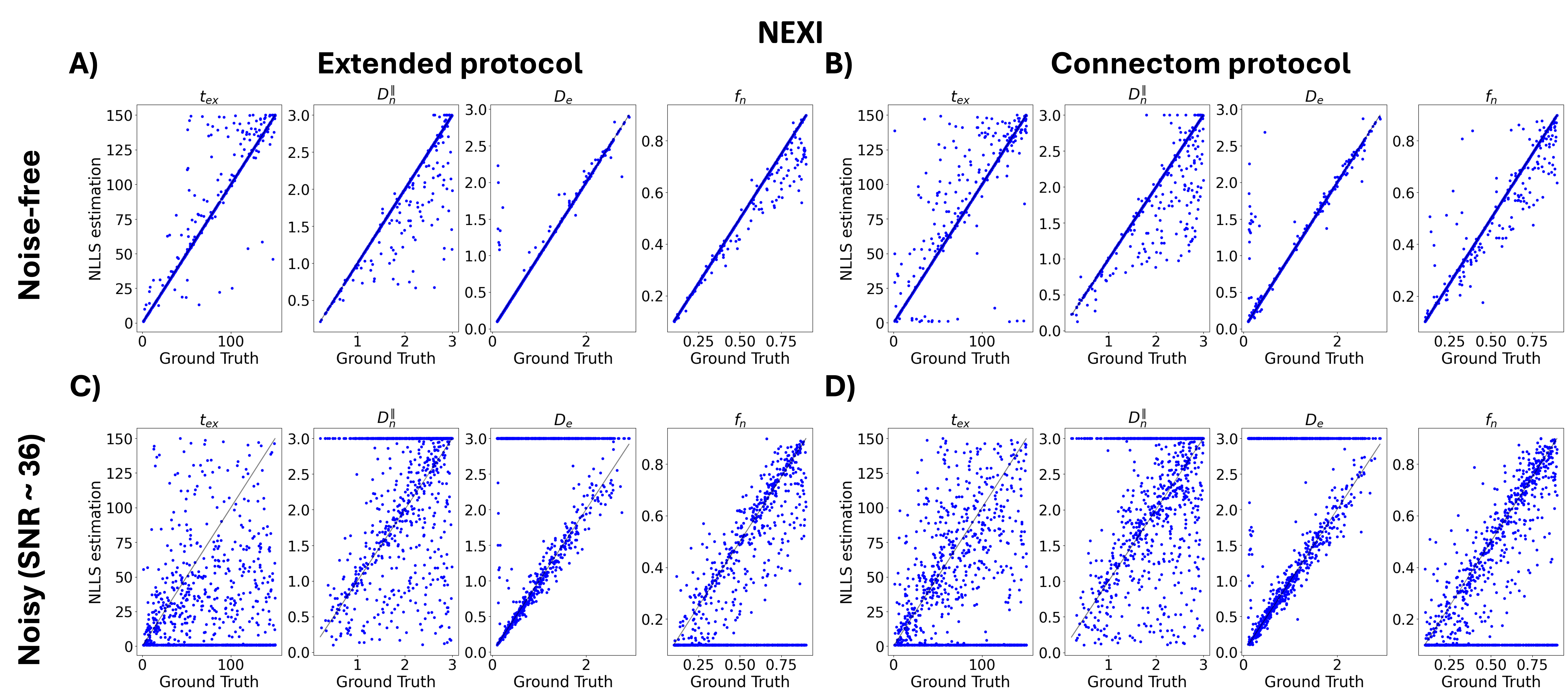}
\caption{Fitting results for the NEXI model using NLLS on 1000 test simulations. Results are shown for the extensive \textit{ex vivo} acquisition protocol (A \& C), and the NEXI 3T Connectom protocol (B \& D). In each subplot, we plot the estimates of the model parameters against their ground truth values.}
\label{fig:suppl_NLLS_fitting_NEXI}
\end{center}
\end{figure}

\begin{figure}[!htbp]
\begin{center}
\includegraphics[width=\textwidth]{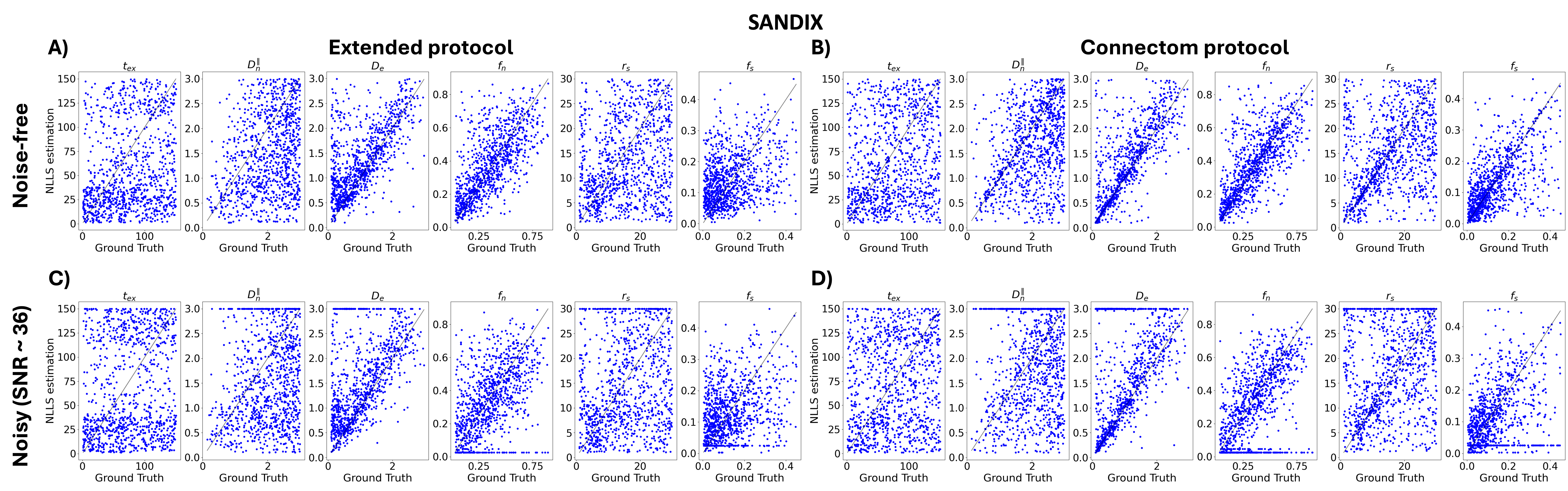}
\caption{Fitting results for the SANDIX model using NLLS on 1000 test simulations. Results are shown for the extensive \textit{ex vivo} acquisition protocol (A \& C), and the NEXI 3T Connectom protocol (B \& D). In each subplot, we plot the estimates of the model parameters against their ground truth values.}
\label{fig:suppl_NLLS_fitting_SANDIX}
\end{center}
\end{figure}

\subsection*{Time Comparison between µGUIDE and NLLS}

\begin{table}[!htbp]
\centering
\caption{Time for fitting 1000 simulations using µGUIDE or NLLS, on both protocols, with and without noise. Time reported excludes the training time for µGUIDE and the initialization for NLLS. Fittings were performed on CPU for both methods (32 cores).
\label{tab:suppl_time_fitting_uGUIDE_NLLS}}
\begin{tabular}{| l || c | *{4}{c|}}
\hline
 \  & \ & \multicolumn{2}{|c|}{Extensive \textit{ex vivo} acquisition protocol} & \multicolumn{2}{|c|}{NEXI 3T Connectom protocol} \\ \cline{3-6}
  \  & \  & µGUIDE & NLLS & µGUIDE & NLLS \\
  \hline
  \multirow{2}{*}{NEXI} & Noise-free & 5s & 45s & 9s & 25s \\ \cline{2-6}
  & Noisy (SNR $\sim$ 50) & 9s & 1min 08s & 9s & 36s \\
  \hline
  \multirow{2}{*}{SANDIX} & Noise-free & 9s & 9min 15s & 7s & 3min 28s \\ \cline{2-6}
  & Noisy (SNR $\sim$ 50) & 7s & 10min 29s & 6s & 3min 22s \\
\hline
\end{tabular}
\end{table}

\subsection*{Scan-rescan Comparison between µGUIDE and NLLS}

\begin{figure}[!htbp]
\begin{center}
\includegraphics[width=0.9\textwidth]{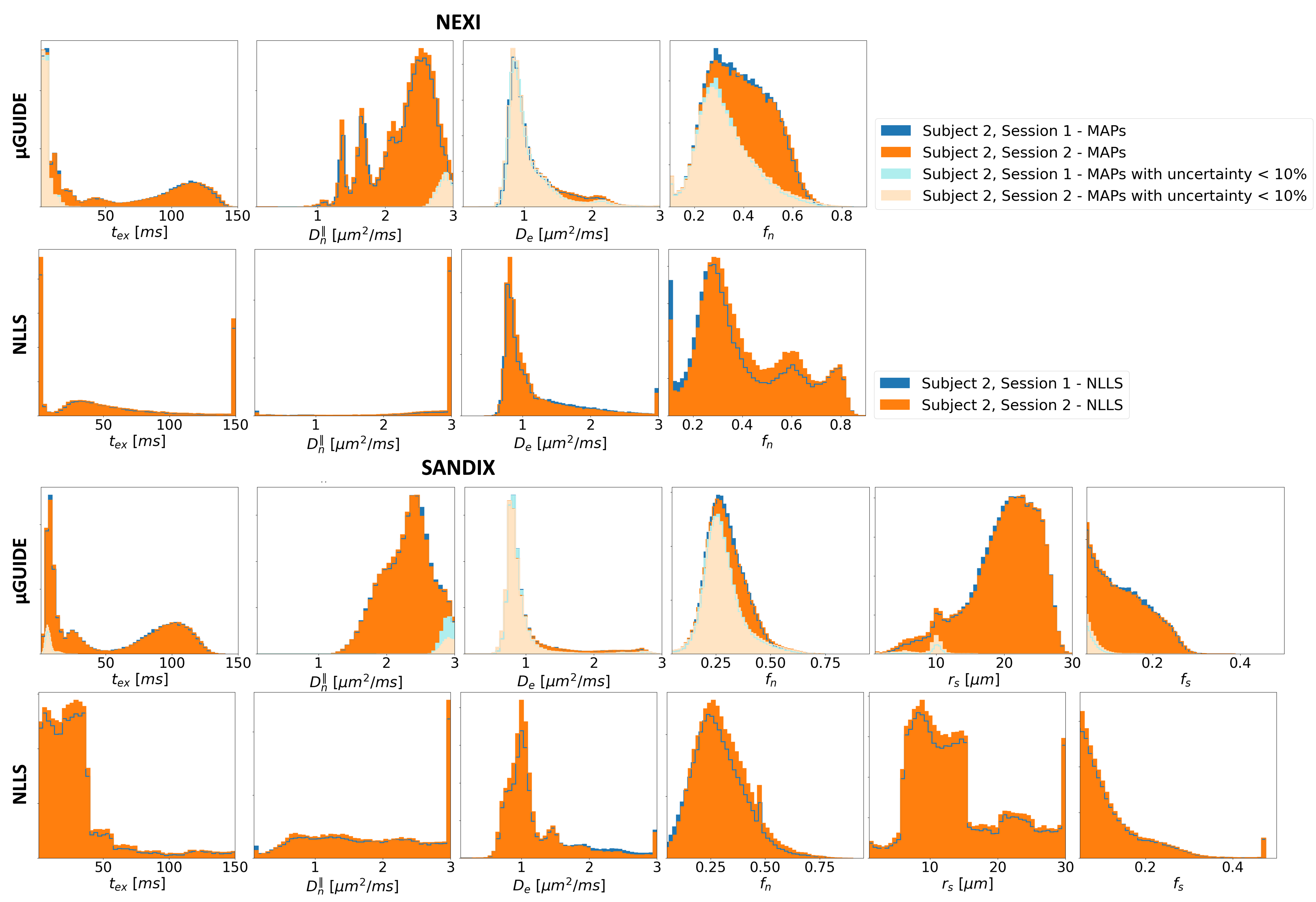}
\caption{Comparison of parameter estimates across cortical ribbon from one participant on two sessions using µGUIDE and a NLLS method. µGUIDE estimates are thresholded based on posterior distribution uncertainty, with distributions shown for voxels with uncertainty below 10\%.}
\label{fig:suppl_uGUIDE_NLLS_scan_rescan}
\end{center}
\end{figure}

\subsection*{Validation of Results Under Gaussian Noise}

\begin{figure}[!htbp]
\begin{center}
\includegraphics[width=0.7\textwidth]{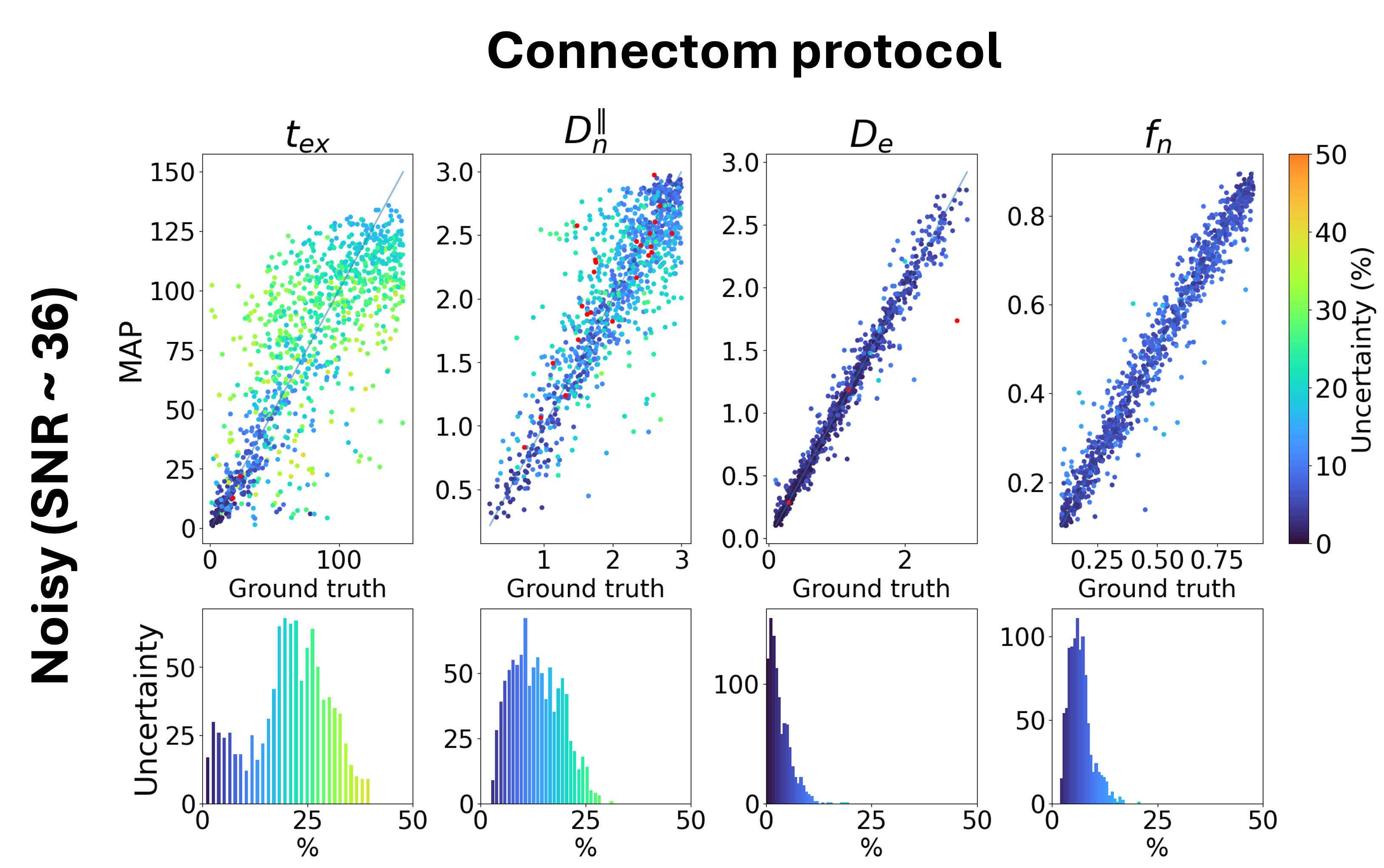}
\caption{Fitting results for the NEXI model using µGUIDE on 1000 simulated signals generated according to the NEXI 3T Connectom protocol and corrupted with Gaussian noise. The top row displays the MAP estimates of the model parameters plotted against their ground truth values, color-coded by their uncertainty values. Red dots indicate cases where the posterior distribution was identified as degenerate. The bottom row shows the distribution of uncertainty values across all test simulations.}
\label{fig:suppl_uGUIDE_NEXI_Gaussian}
\end{center}
\end{figure}

\printbibliography

\end{document}